\documentclass[12pt]{article}
\usepackage{bbm}
\usepackage{fullpage}
\usepackage{float}
\usepackage{amsfonts,amsmath,amssymb}
\usepackage{mathtools}
\usepackage{mathrsfs}
\usepackage{dsfont}
\usepackage{color}
\usepackage{graphicx}
\usepackage{url}

\usepackage{algorithm}
\usepackage{algpseudocode}

\def\defi{\vcentcolon=}

\newcommand{\ignore}[1]{}

\newtheorem{example}{Example}[section]

\author{}
\date{}

\title{Management and Visualization Tools for Emergency Medical Services}

\begin{document}

\maketitle
\vspace*{-2cm}

\begin{center}
\begin{tabular}{ccc}
\begin{tabular}{c}
Vincent Guigues\\
School of Applied Mathematics, FGV\\
Praia de Botafogo, Rio de Janeiro, Brazil\\
{\tt vincent.guigues@fgv.br}
\end{tabular}&
&
\begin{tabular}{c}
Anton J. Kleywegt\\
Georgia Institute of Technology\\
Atlanta, Georgia 30332-0205, USA\\
{\tt anton@isye.gatech.ed}\\
\end{tabular}\\
&&\\
\begin{tabular}{c}
Victor Hugo Nascimento\\
School of Applied Mathematics, FGV\\
Praia de Botafogo, Rio de Janeiro, Brazil\\
{\tt victorhugo.vhrn@gmail.com}
\end{tabular}
&
&
\begin{tabular}{c}
Victor Salles Rodrigues\\
Grama\\
Rio de Janeiro, Brazil\\
{\tt victorsalles@me.com}
\end{tabular}\\
&&\\
\begin{tabular}{c}
Thais Viana\\
Grama\\
Rio de Janeiro, Brazil\\
{\tt thaisnviana@gmail.com }
\end{tabular}
&
&
\begin{tabular}{c}
Edson Medeiros\\
Grama\\
Rio de Janeiro, Brazil\\
e@edsonmedeiros.com
\end{tabular}
\end{tabular}
\end{center}

\begin{abstract}
This paper describes an online tool for the visualization of medical emergency locations, randomly generated sample paths of medical emergencies, and the animation of ambulance movements under the control of various dispatch methods in response to these emergencies.
The tool incorporates statistical models for forecasting emergency locations and call arrival times, the simulation of emergency arrivals and ambulance movement trajectories, and the computation and visualization of performance metrics such as ambulance response time distributions.
Data for the Rio de Janeiro Emergency Medical Service are available on the website.
A user can upload emergency data for any Emergency Medical Service, and can then use the visualization tool to explore the uploaded data.
A user can also use the statistical tools and/or the simulation tool with any of the dispatch methods provided, and can then use the visualization tool to explore the computational output.
Future enhancements include the ability of a user to embed additional dispatch algorithms into the simulation; the tool can then be used to visualize the simulation results obtained with the newly embedded algorithms.
\end{abstract}

\section{Introduction}

The management of an emergency medical service (EMS) is a complex process that includes the real-time allocation of ambulances and crews to emergency calls received at a call center.
Tools for the visualization of emergency data, forecasts, ambulance movements, and performance metrics, are of great benefit in the planning and real-time control of emergency services, the training of personnel, and the development of management tools.
In this paper, we describe a visualization tool that we developed for the efficient management of an EMS.
The tool is available at \url{http://samu.vincentguigues.com}.
The tool also includes statistical methods for forecasting medical emergencies described in \cite{laspatedpaper} and \cite{laspatedmanual}, available at \url{https://github.com/vguigues/LASPATED}, as well as ambulance dispatch algorithms described in \cite{ourheuristics23a}, and available at \url{https://github.com/vguigues/Heuristics_Dynamic_Ambulance_Management}.

\paragraph{Ambulance operations.}
Next, we give an overview of the events that happen during the response to a medical emergency.
When an emergency call arrives at a call center, the call is assigned to a call taker, also called a telecommunicator or an emergency medical dispatcher.
The call taker asks the caller a sequence of questions, and the answers are used to classify the emergency.
There are various classification systems such as the International Classification of Diseases (ICD), the Medical Priority Dispatch System (MPDS), and the Association of Public-Safety Communications Officials (APCO) system, and many EMSs have their own systems.
In outline, emergencies are classified based on the body part affected and/or the cause of the emergency, as well as the level of urgency.
Based on the classification, the call taker decides whether an ambulance is dispatched immediately to the call, and if so, which ambulance and crew are dispatched, or whether the call is put in a queue of calls for which an ambulance will be dispatched later.
This decision problem is called the \emph{ambulance selection} problem.
When an ambulance arrives at the scene of an emergency, the crew provides assistance as they deem appropriate.
The service may be completed at the scene of the emergency, or one or more patients may be transported to a hospital.
After handing patients over to the hospital, the ambulance has to be cleaned.
Depending on the situation, the cleaning task may be quick and simple and can be completed before departing the hospital, or may require the ambulance to travel to a cleaning station for more thorough cleaning.
After an ambulance has finished service, at the scene of the emergency or at a hospital or at a cleaning station, a decision is made where to send the ambulance next: either to a call in queue, and if so to which call in queue, to an ambulance station, where the ambulance waits while not in service.
This decision problem is called the \emph{ambulance reassignment} problem.

EMS managers use various performance metrics.
One of the main performance metrics is response time.
There are a number of response time metrics, such as the amount of time elapsed from the instant the call was received to the instant the first emergency crew arrives at the emergency site, or the amount of time elapsed from the instant the call was received to the instant the patient is handed over to the hospital.
The impact of response time on patient outcome depends on the type of emergency and other factors that are often not well understood \cite{cret:79,blac:91,lars:93,vale:97,stie:99,vale:00,waal:01,pell:01,dema:03,pons:05,blac:09,blan:12,weis:13}.
One of the important factors is the capabilities of the ambulance and crew.
There are different types of ambulances, such as Basic Life Support (BLS), Intermediate Life Support (ILS), and Advanced Life Support (ALS) ambulances, as well as Mobile Stroke Units (MSU), ambulance helicopters, etc.
The emergency medical personnel also have different amounts of experience and different qualifications, such as Emergency Medical Technician (EMT), Advanced Emergency Medical Technician (AEMT), paramedic, and physician.
The type of emergency as well as the capabilities of the ambulance and crew should be considered when making ambulance dispatch decisions.

\paragraph{Related literature.}
Research on ambulance management has addressed various planning and operational problems.
Many papers have addressed the location of ambulance stations or the assignment of ambulances to stations, including \cite{tore:71,berl:74,chur:74,schi:79,dask:81,dask:83,hoga:86,reve:89,repe:94,gend:97,ingo:08,erku:09,schm:10}, and \cite{sore:10}.
Various stochastic models, including queuing models and simulations, have been proposed by \cite{volz:71,swov:73a,swov:73b,fitz:73,lars:74,lars:75,hill:84,jarv:85,gold:90b,gold:91a,gold:91b,burw:93} and \cite{rest:09} for evaluating location decisions for stations and ambulances.
The most popular policy in the literature for the ambulance selection problem is the simple closest-available-ambulance rule, used by \cite{hend:99,hend:04,maxw:09,maxw:10,maxw:13}, and \cite{alan:13}.
As the name indicates, when an emergency call arrives, the closest-available-ambulance rule dispatches the available ambulance that is closest (in terms of forecasted response time) to the emergency.
A few papers have proposed alternatives to the closest-available-ambulance rule, including
\cite{ande:07,lees:11,schm:12,band:12,mayo:13,band:14,lisay:16,jagt:17a} and \cite{jagt:17b}.
Fewer papers have addressed the ambulance reassignment decision (where to send an ambulance when it becomes available).
One popular approach is the following:
First, the ambulance stations are located.
Second, each ambulance is assigned to a station, called its home station.
Then, during operations, when an ambulance becomes available and is not dispatched to an emergency waiting in queue, the ambulance is sent to its home station \cite{gold:90b,hend:04,rest:09,band:12,knig:12,maso:13,mayo:13,band:14}.
In \cite{guiklevhn2022}, a stochastic optimization model is proposed for both the ambulance selection problem and the ambulance reassignment problem.

\paragraph{Contributions.}
Below we describe the main functionalities of the website:
\begin{itemize}
\item[(1)]
\textbf{Repository of EMS data.}
One of the goals of this website is to serve as a repository of EMS data and benchmark instances of EMS problems that can be used by EMSs for training and by researchers working on EMS problems.
The website provides data provided by the Rio de Janeiro EMS.
These data can be downloaded and include a $2$~year history of emergency calls, the locations of ambulance stations, the locations of hospitals, and the set of ambulances and their types.
Template files for formatting the same types of data are given.
\item[(2)]
\textbf{Visualization of EMS data.}
As described in Section~\ref{sec:visualization}, the website offers several visualization tools (line plots, heatmaps, histograms, pie charts) for the visualization of emergency calls.
In addition, EMS data such as the locations of ambulance stations and the locations of hospitals are visualized on a map.
\item[(3)]
\textbf{Visualization of randomly generated sample paths of emergencies.}
Statistical methods for forecasting medical emergencies are described in \cite{laspatedpaper} and \cite{laspatedmanual}, and are available at \url{https://github.com/vguigues/LASPATED}.
These methods are also provided on the website.
In addition, as described in Section~\ref{sec:forecast visualization}, the website provides tools for the visualization of randomly generated sample paths in the form of heatmaps, histograms, and line plots.

\ignore{
\begin{figure}
    \centering
    \includegraphics[width=0.67\linewidth]{disc_10x10.pdf}
    \caption{Discretization of the city of Rio de Janeiro in 76 rectangular regions.}
    \label{label:disc_rect}
\end{figure}

\begin{figure}
    \centering
    \includegraphics[width=0.67\linewidth]{disc_hex7.pdf}
    \caption{Discretization of the city of Rio de Janeiro in 226 hexagonal regions.}
    \label{label:disc_hex}
\end{figure}

\begin{figure}
    \centering
    \includegraphics[width=0.67\linewidth]{disc_custom.pdf}
    \caption{Discretization of the city of Rio de Janeiro into its 160 administrative districts.}
    \label{label:disc_district}
\end{figure}
}

\item[(4)]
\textbf{Simulation of ambulance operations.}
The website provides a tool to simulate ambulance operations in any service region of interest.
The simulation mimics the ambulance operations described above.
More details of the simulation model are given in Section~\ref{sec:ambmodel}.
The user can choose several parameters such as the duration of the simulation run and the ambulance dispatch policy.
The simulation output can be downloaded.
\item[(5)]
\textbf{Visualization of ambulance operations and performance metrics.}
The user can use the website to visualize the simulated ambulance operations, including the arrival of emergencies, the current location and movement of ambulances, and the delivery of patients to hospitals.
Given a graph of the streets, the visualization shows the movement of ambulances along the streets, and not along the great circle path between the origin point and the destination point.
The user can also use the website to visualize performance metrics obtained from the simulation output, such as the distribution of ambulance response times.
\end{itemize}
This website could be adapted to provide outputs for other settings involving vehicle movements, such as police and fire response management, together with associated data and benchmark instances, and dispatch policies.

\paragraph{Organization of the paper.}
Section~\ref{sec:visualization} describes the functionalities for visualizing emergency data.
Section~\ref{sec:forecast visualization} describes the forecast tools, and the forecast visualization functionalities.
Section~\ref{sec:ambmodel} describes the simulation and visualization of ambulance operations, including the computation and visualization of ambulance trajectories as well as performance metrics under specified ambulance dispatch policies.
Section~\ref{sec:tech_stack} briefly addresses the software technologies chosen for this project.
Concluding remarks are given in Section~\ref{sec:conclusion}.
All figures were produced using the website.

\section{Visualization of Emergency Calls}
\label{sec:visualization}

The visualization and forecast of emergency calls are based on emergency call data.
The data are stored in a table with one row for each emergency, and various fields, including date of the call, time of the call, emergency type, (latitude,longitude) location of the emergency, ambulance arrival time at the emergency, hospital (if applicable), and ambulance arrival time at the hospital (if applicable).
The emergency type can be according to any classification system, and can be the emergency type as classified by the call taker, or the emergency type as classified by the ambulance personnel in their emergency report.

The link {\em{Visualization of emergency calls}} 
accessible from the link
{\em{SAMU - Emergency calls}} allows us to access a webpage
with five buttons on top: an {\em{Upload data}} button to upload new data,   and four buttons 
{\em{line plot}}, {\em{heatmap}}, {\em{histogram}}, and {\em{pie chart}}, see Figure \ref{fig2}. We illustrate below these five
visualization tools.

\subsection{Upload of new data}
\label{sec:upload}

The first step to visualize data of emergency calls is to select a  database of emergency calls
in the webpage  upper-left corner filter {\em{Database}}. 
The current website contains a database of Rio de Janeiro EMS emergency calls, that is available to all users.

\begin{figure}
\centering
\begin{tabular}{c}
\includegraphics[scale=0.18]{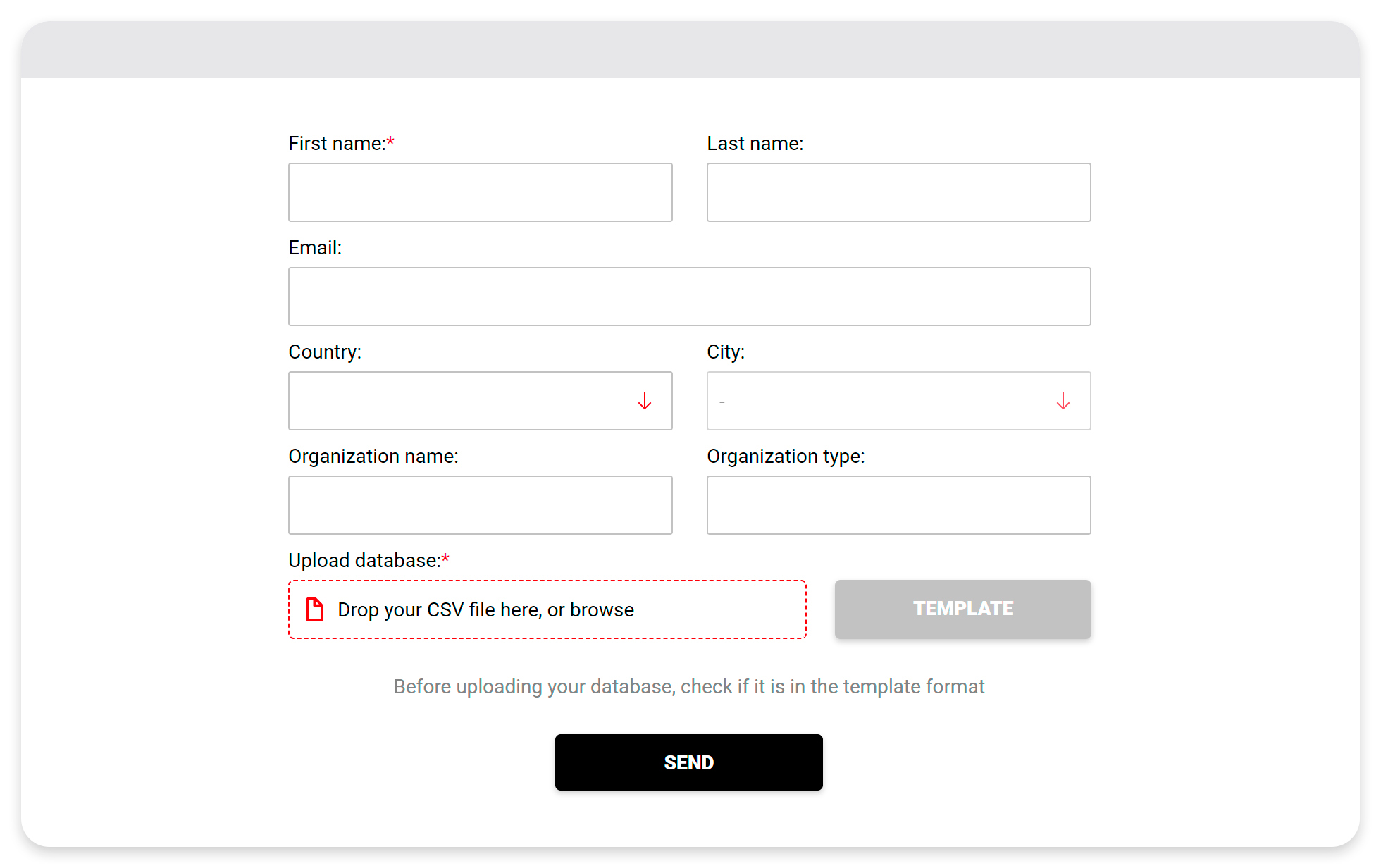}
\end{tabular}
\caption{Form to upload new emergency call data.}\label{fig1}
\end{figure}

Historical emergency call data can be uploaded using the form shown in Figure~\ref{fig1}.
The data are checked for validity before enabling the data to be used with other functionalities of the tool.

Four methods to visualize emergency data are provided clicking the aforementioned four buttons: line plot, heatmap, histogram, and pie chart.

\subsection{Line plots}

\begin{figure}
\centering
\begin{tabular}{c}
\includegraphics[scale=0.18]{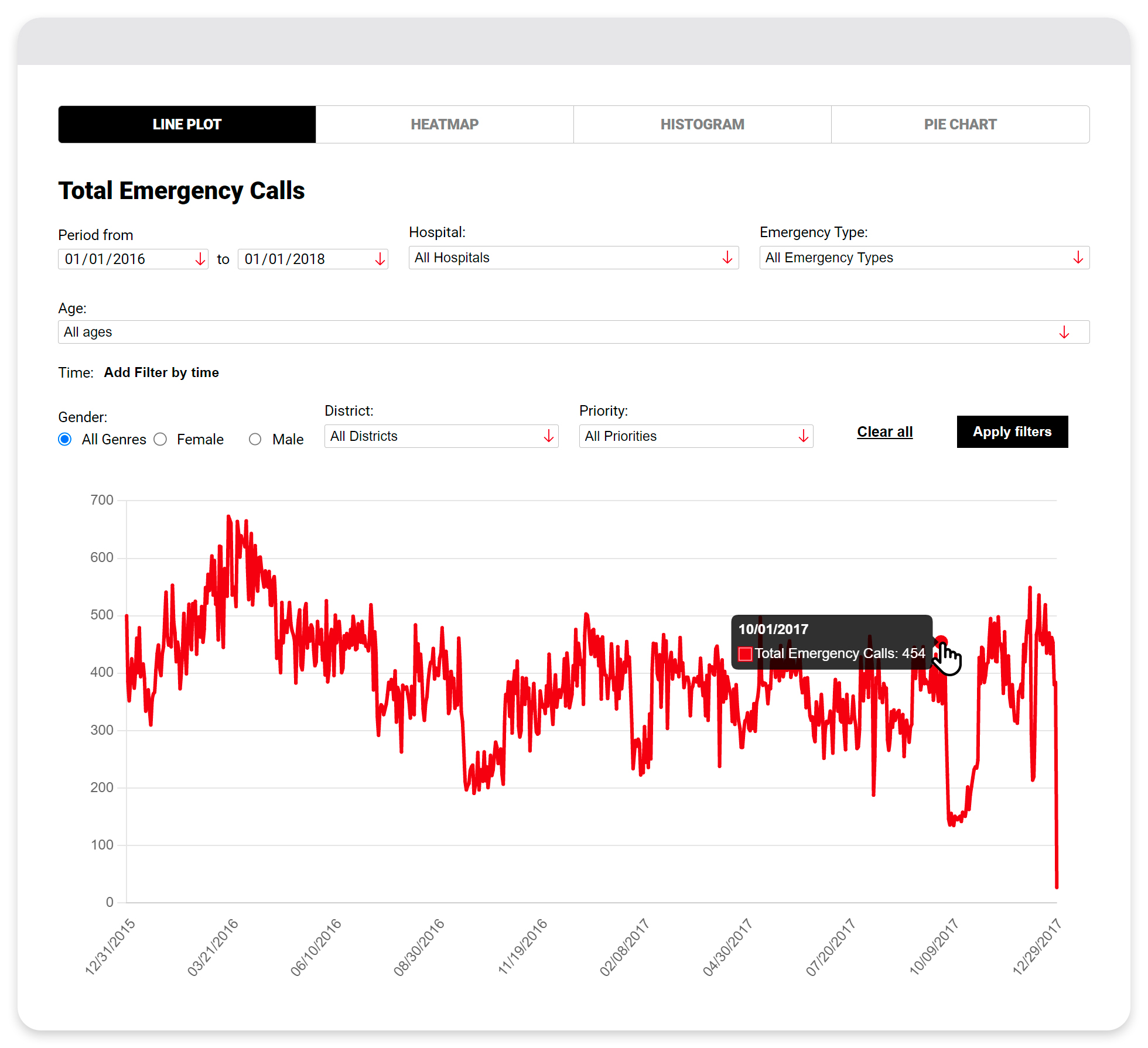}
\end{tabular}
\caption{Line plot of the empirical rate of emergencies as a function of time.}\label{fig2}
\end{figure}

It is important for the management of emergency medical services to be familiar with the rates at which emergencies take place, and how the rates vary over time.
A line plot can be used to visualize how the rate at which emergencies take place vary over time, such as by time of day, day of week, etc. Line plots can be
obtained clicking 
link
{\em{Visualization of emergency calls}}
then clicking the {\em{Line plot}}
button. The tool uses the emergency data to compute empirical rates, and displays the empirical rates as shown in Figure~\ref{fig2}.
The download data button selects the database of emergency calls used for the calculations and visualization.
The following filters can be used to control the subset of data used to plot the empirical rate of emergencies as a function of time:
(1)~the time period of data used (beginning date, end date),
(2)~the time windows during the day (all or selected time windows of 30 mins),
(3)~the emergency type(s),
(4)~the age groups of the patients,
(5)~the gender of the patients,
(6)~the hospitals (all or specific hospitals),
(7)~the EMS districts (all or specific districts),
(8)~the priority levels of the emergencies.

\subsection{Heatmaps}

\begin{figure}[H]
\centering
\begin{tabular}{c}
\includegraphics[scale=0.17]{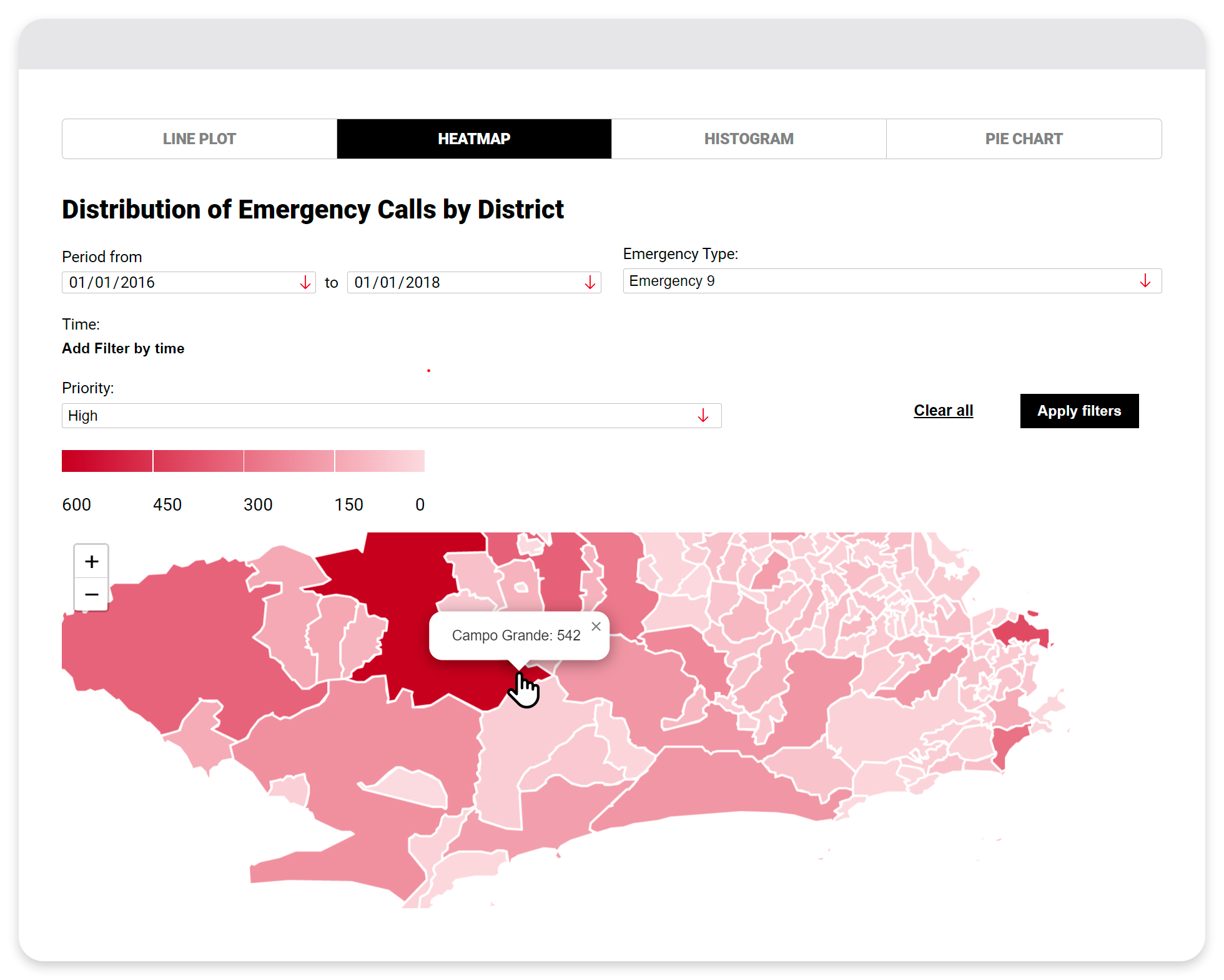}
\end{tabular}
\caption{Heatmap of the empirical intensity of emergencies as a function of space.}\label{fig3}
\end{figure}

Whereas line plots are used to visualize the empirical rate of emergencies as a function of time, heatmaps are used to visualize the empirical rate of emergencies as a function of space, as shown in Figure~\ref{fig3}.
Heatmaps can be obtained clicking the
link
{\em{Visualization of emergency calls}}
then clicking
the {\em{Heatmap}} button.
The following filters can be used to control the subset of data used to plot the empirical intensity of emergencies as a function of space:
(1)~the time period of data used (beginning date, end date),
(2)~the time windows during the day (all or selected time windows of 30 mins),
(3)~the emergency type(s),
(4)~the priority levels of the emergencies.

\subsection{Histograms}

\begin{figure}
\centering
\begin{tabular}{c}
\includegraphics[scale=0.17]{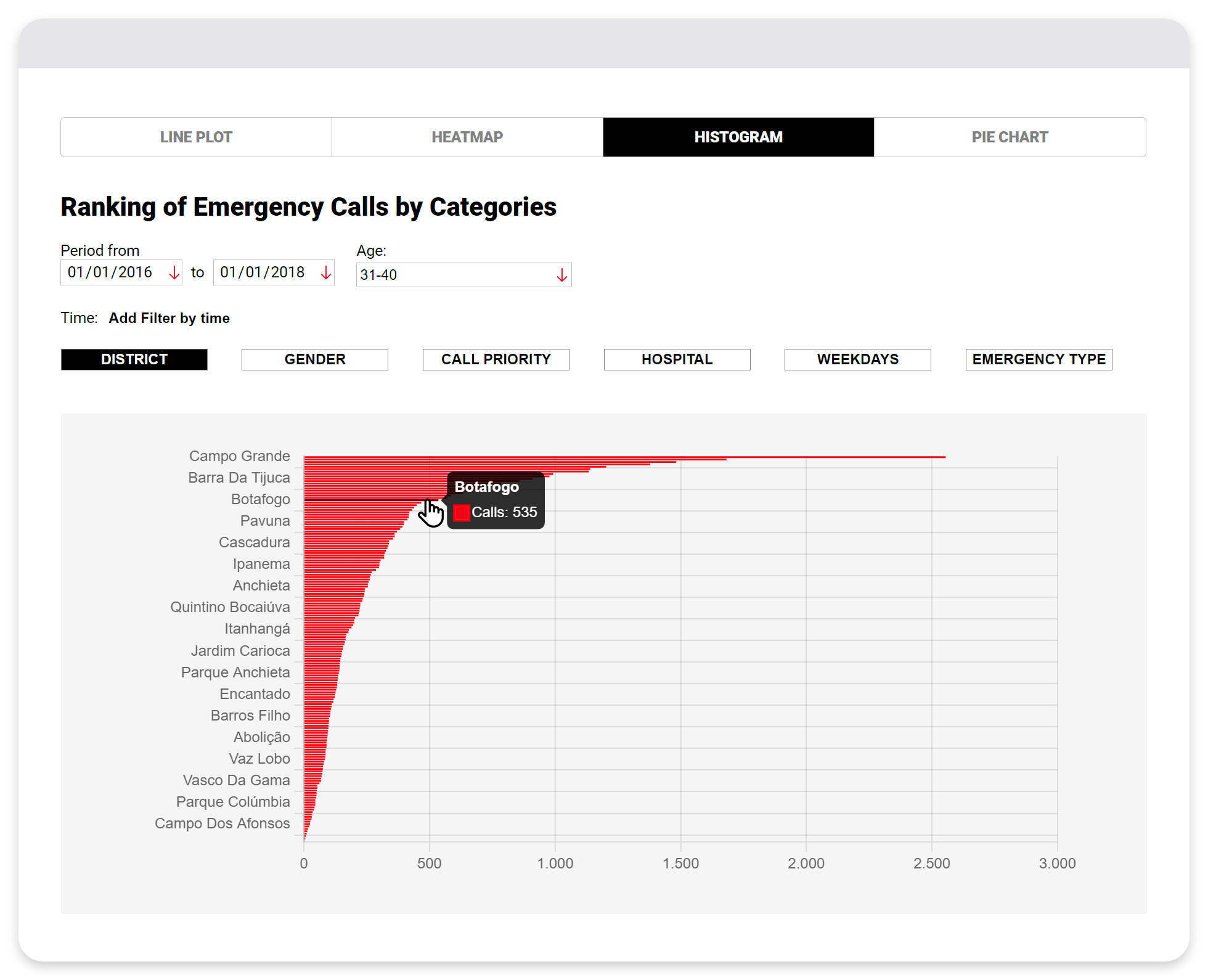}
\end{tabular}
\caption{Histogram showing the ranking of empirical emergency rates.}\label{fig4}
\end{figure}

Histograms are used to visualize the ranking of the empirical rates of emergencies by one of the following categorical variables: day of the week, emergency type, priority level, district, gender, and hospital.
Histograms can be obtained clicking
link
{\em{Visualization of emergency calls}}
then clicking
the {\em{Histogram}} button.
An example histogram showing the ranking by district is shown in Figure~\ref{fig4}.
The following filters can be used to control the subset of data used to plot the histogram of empirical emergency rates:
(1)~the time period of data used (beginning date, end date),
(2)~the time windows during the day (all or selected time windows of 30 mins),
(3)~the age groups of the patients.

\subsection{Pie charts}

\begin{figure}
\centering
\begin{tabular}{c}
\includegraphics[scale=0.19]{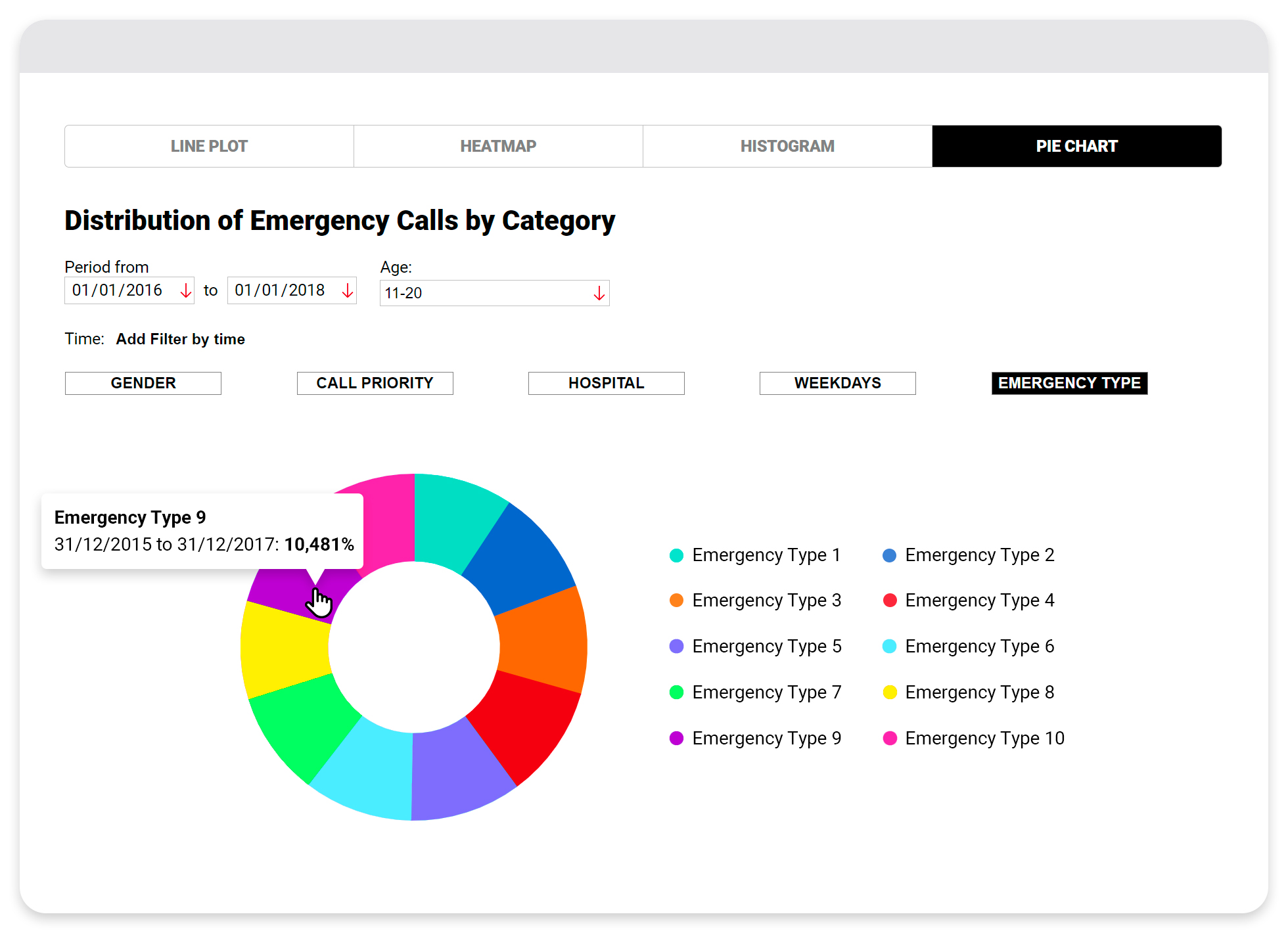}
\end{tabular}
\caption{Empirical distribution of emergencies by emergency type.}\label{fig5}
\end{figure}

Pie charts are used to visualize the fractions of emergencies by one of the following categorical variables: day of the week, emergency type, priority level, gender, and hospital.
Pie charts can be obtained clicking
link
{\em{Visualization of emergency calls}}
then clicking
the {\em{Pie chart}} button.
An example pie chart showing the fractions of emergencies by emergency type is shown in Figure \ref{fig5}.
The following filters can be used to control the subset of data used to plot the pie chart of empirical emergency rates:
(1)~the time period of data used (beginning date, end date),
(2)~the time windows during the day (all or selected time windows of 30 mins),
(3)~the age groups of the patients.

\section{Visualization of Randomly Generated Arrivals of Emergencies}
\label{sec:forecast visualization}

The website provides an interface for the statistical methods for forecasting medical emergencies as described in \cite{laspatedpaper} and \cite{laspatedmanual}.
The website also provides tools for the visualization of randomly generated arrivals of emergencies in the form of heatmaps, histograms, and line plots.
The link
{\em{Emergency calls forecast}} allows us to access
these tools.

Two filters must first be selected.

The first filter {\em{Model}} requires to select a model. 
To select the model without
covariates briefly described
in Section \ref{sec:forecastcall} (see details of this model in \cite{laspatedpaper}, \cite{laspatedmanual}),
select option {\em{No regression}}. To select the model with
covariates briefly described
in Section \ref{sec:forecastcall} (see details of this model in \cite{laspatedpaper}, \cite{laspatedmanual}),
select option {\em{Regression}}.

The second filter {\em{Discretization type}} provides the discretization scheme. For the model without covariates, the allowed discretizations are rectangular 10x10 (option Rect 10), hexagonal discretization  obtained with Uber library h3 and scale parameter equal to 7 (option Uber 7), and by districts (option District). For the model with covariates, the allowed discretizations are rectangular 10x10 and by districts. We then have three selectors: heatmap, histogram, and line plot.

\subsection{Forecast of emergency calls}
\label{sec:forecastcall}

Here we give a brief overview of the statistical methods for forecasting medical emergencies described in \cite{laspatedpaper} and \cite{laspatedmanual}.
The methods are based on discretization of space and time.
Several spatial discretization methods are available for use, such as partitioning into rectangles, hexagons, partitioning by district, or custom partitioning.
Similarly, several time discretization methods are available for use.
Let $\mathcal{I}$ denote the subsets of the space discretization, and let $\mathcal{T}$ denote the subsets of the time discretization.
Let $\mathcal{C}$ denote the set of emergency types.
Provision is made for two types of models:
\begin{enumerate}
\item Model without covariates. 
Emergency arrival rate $\lambda_{c,i,t}$ for each emergency type $c \in \mathcal{C}$, zone $i \in \mathcal{I}$, and time period $t \in \mathcal{T}$.
This model does not use covariates.
Model parameters $\lambda_{c,i,t}$ are estimated via maximum likelihood estimation.
Regularization of the estimation problem can be done by specifying neighboring zone pairs $(i,j) \in \mathcal{I}^2$ and neighboring time pairs $(t,t') \in \mathcal{T}^2$, in which case the regularization weights are determined by cross-validation.
\item Model with covariates.
Emergency arrival intensity $\lambda(x) = \beta^{\top} x$ as a function of a vector $x = (x_{1},\ldots,x_{K})$ of covariate values.
Typically, the covariate values depend on the emergency type $c \in \mathcal{C}$, zone $i \in \mathcal{I}$, and time period $t \in \mathcal{T}$.
The model parameters $\beta = (\beta_{1},\ldots,\beta_{K})$ are estimated via constrained maximum likelihood estimation.
By choosing each covariate $x_{k}$ to be an indicator of a specific combination of emergency type $c \in \mathcal{C}$, zone $i \in \mathcal{I}$, and time period $t \in \mathcal{T}$, it can be seen that the model without covariates is a special case of the model with covariates (with $\beta_{k} = \lambda_{c,i,t}$).
However, this special case is much easier to specify directly as above rather than through cumbersome covariates.
\end{enumerate}

\subsection{Line plots}

\begin{figure}
\centering
\begin{tabular}{c}
\includegraphics[scale=0.25]{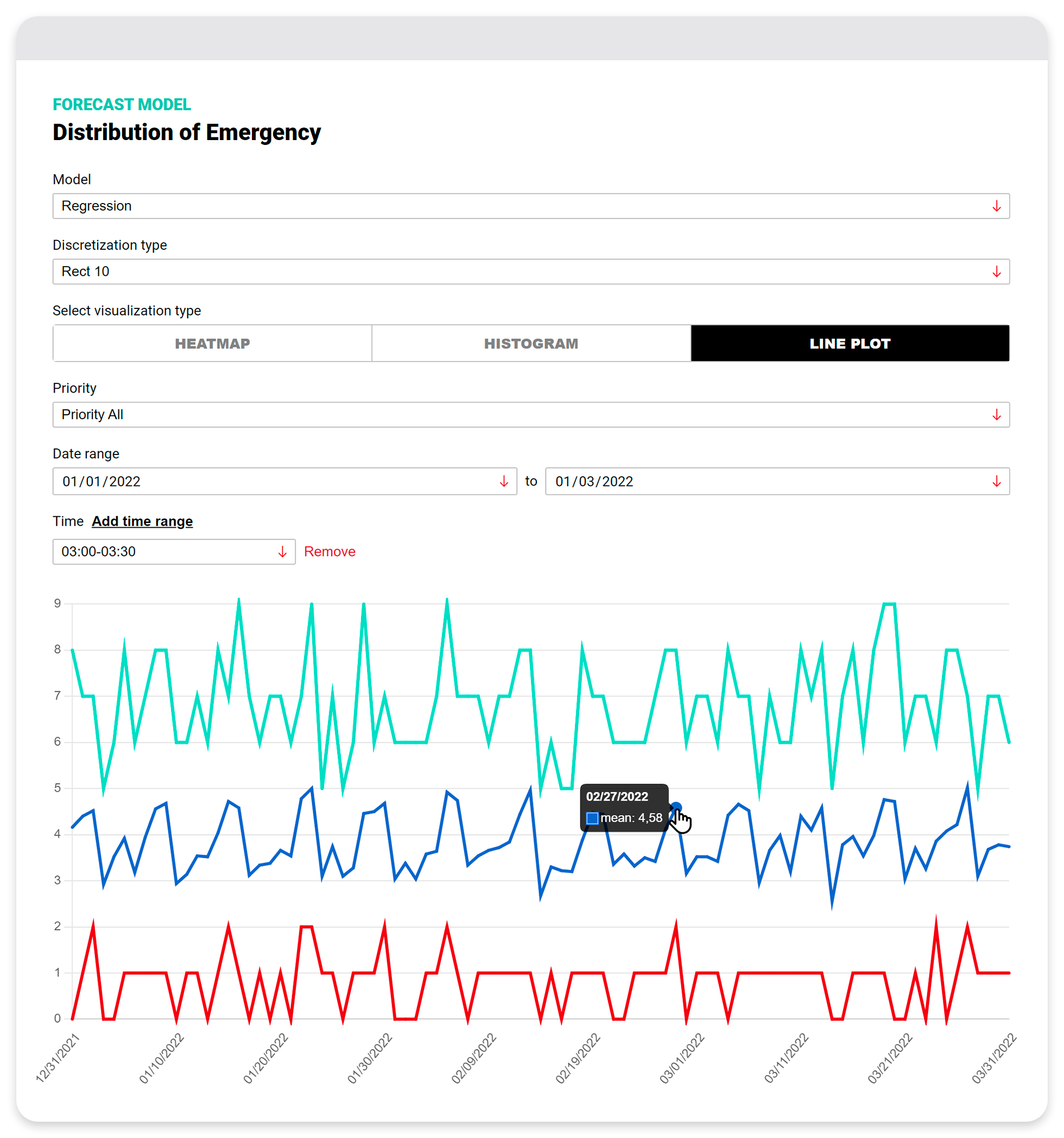}
\end{tabular}
\caption{Mean, and $0.05$ and $0.95$ quantiles of the randomly generated number of emergencies.}\label{fig8}
\end{figure}

Line plots are used to visualize randomly generated numbers of emergency arrivals as a function of time.
They are generated clicking
the {\em{Line plot}} button.
The following filters can be used to control the subset of emergencies for which the numbers of arrivals as a function of time is shown:
(1)~the date range (beginning date, end date),
(2)~the time windows during the day (all or selected time windows of 30 mins),
(3)~the priority levels of the emergencies.
Figure~\ref{fig8} shows an example of such a line plot.
Note that for each chosen time period it shows the forecasted mean number of emergencies,
as well as the $0.05$ and $0.95$ quantiles of the generated number of emergencies.

\subsection{Heatmaps}

\begin{figure}
\centering
\begin{tabular}{c}
\includegraphics[scale=0.25]{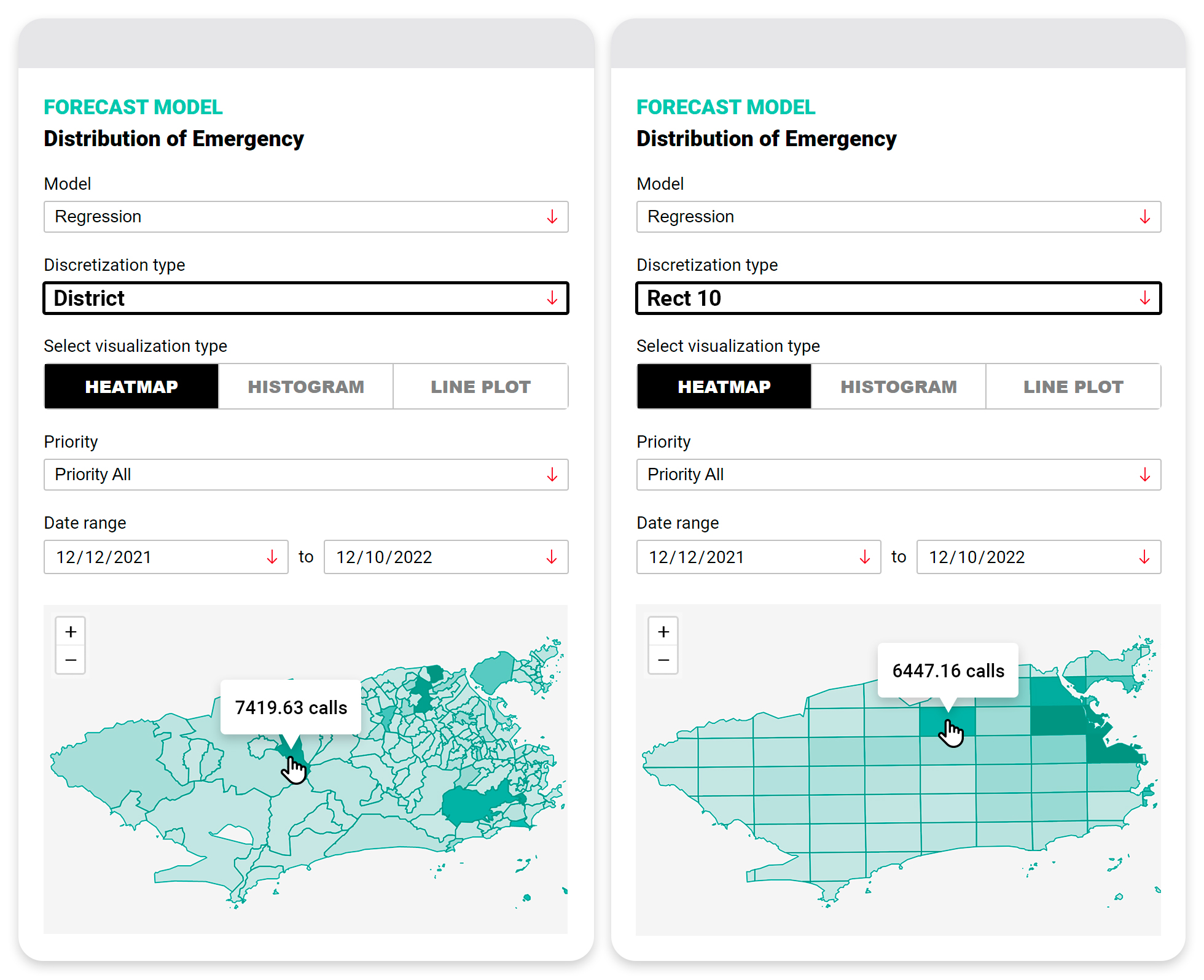}
\end{tabular}
\caption{Heatmap of forecasted number of emergencies.}\label{fig6}
\end{figure}

Heatmaps are used to visualize the forecasted number of emergencies as a function of space.
They are generated clicking
the {\em{Heatmap}} button.
The following filters can be used to control the subset of emergencies for which the generated number of arrivals as a function of space is shown:
(1)~the date range (beginning date, end date), 
(2)~the time windows during the day (all or selected time windows of 30 mins),
(3)~the priority levels of the emergencies.
Figure~\ref{fig6} shows an example of such heatmaps for two space discretization methods.

\ignore{
\subsection{Histograms}

Histograms are used to visualize the distribution of the randomly generated number of emergencies.
The following filters can be used to control the subset of emergencies for which the distribution of the generated number of arrivals is shown:
(1)~the date range (beginning date, end date),
(2)~the time windows during the day (all or selected time windows of 30 mins),
(3)~the priority levels of the emergencies.
An example of such a histogram is shown in Figure~\ref{fig7}.

\begin{figure}
\centering
\begin{tabular}{c}
\includegraphics[scale=0.25]{histogramforecast.png}
\end{tabular}
\caption{Histogram of the distribution of the randomly generated number of emergencies.}\label{fig7}
\end{figure}
}

\section{Simulation and Visualization of Ambulance Operations}
\label{sec:ambmodel}

An overview of ambulance operations was given in the introduction.
In this section we describe a simulation and visualization of ambulance operations that can be accessed through the website.

\subsection{Simulation of ambulance operations}
\label{sec:ambfleet}

In this section, we describe how the simulation keeps track of the state of ambulance operations, and how ambulance selection and ambulance reassignment decisions are incorporated into the simulation.
The source code of the simulation is available at \url{https://github.com/vguigues/Heuristics_Dynamic_Ambulance_Management}.
The variable and parameter names below match the names used in the source code.
The following variables are associated with each emergency, indexed with~$i$:
\begin{itemize}
\item
the time instant $t_{c}(i)$ of the emergency call;
\item
the location $\ell_{c}(i)$ of the emergency, given by a (latitude,longitude) pair;
\item
the type $\mbox{Type}_{c}(i)$ of the emergency;
\item
the amount of time $\mbox{WaitingOnScene}(i)$ between the instant of the emergency call and the instant that an ambulance arrives on the scene of the emergency;
\item
the penalty $\mbox{WaitingOnScenePenalized}(i)$ for the amount of time between the instant of the emergency call and the instant that an ambulance arrives on the scene of the emergency: the penalized waiting time is $$
\mbox{WaitingOnScenePenalized}(i) \\ = {\tt penalization}(\mbox{WaitingOnScene}(i),\mbox{Type}_{c}(i))
$$ for some penalization function $\tt{penalization}$, see also \eqref{costalloc};
\item
the amount of time $\mbox{TimeOnScene}(i)$ spent by the ambulance on the scene of the emergency
(the time between the instant an ambulance arrives on the scene of emergency~$i$ and the instant the ambulance departs from the scene of emergency~$i$);
\item
the hospital location $\ell_{h}(i)$ where the patient(s) are taken, if applicable;
\item
the amount of time $\mbox{WaitingToHospital}(i)$ between the instant that the ambulance arrives on the scene of the emergency and the instant that the ambulance arrives at the hospital, if applicable;
\item
the penalty $\mbox{WaitingToHospitalPenalized}(i)$
is given by
$${\tt penalization}(\mbox{WaitingToHospital}(i),\mbox{Type}_{c}(i))$$ for the amount of time between the instant that the ambulance arrives on the scene of the emergency and the instant that the ambulance arrives at the hospital, if applicable;
\item
the amount of time $\mbox{TimeAtHospital}(i)$ after transporting the patient that the ambulance waits for the patient(s) to be admitted to the hospital, if applicable;
\item
the cleaning station location $\ell_{cb}(i)$ where the ambulance is cleaned after serving emergency~$i$, if applicable;
\item
the time $\mbox{CleaningTime}(i)$ that the ambulance spends at the cleaning station after serving emergency~$i$, if applicable;
\item
the ambulance station location $B(i)$ where the ambulance is sent after serving emergency~$i$, if applicable;
\end{itemize}
The following variables are associated with each ambulance, indexed with indexAmb:
\begin{itemize}
\item
the type $\mbox{Type}_{a}(\mbox{indexAmb})$ of the ambulance;
\item
if ambulance indexAmb is currently serving an emergency, then $t_{f}(\mbox{indexAmb})$ is the time that the ambulance is scheduled to complete the service, and if ambulance indexAmb is currently on its way to a station or at a station, then $t_{f}(\mbox{indexAmb})$ is the time when the ambulance completed the previous service;
\item
if ambulance indexAmb is currently serving an emergency, then $\ell_{f}(\mbox{indexAmb})$ is the location where service of the emergency will be completed (the location of the hospital if patient(s) are transported to a hospital and the ambulance is cleaned at the hospital, or the location of the cleaning station if the ambulance is cleaned at a cleaning station, otherwise the location of the emergency), given by a (latitude,longitude) pair, otherwise $\ell_{f}(\mbox{indexAmb})$ is the location where service of the previous emergency served by ambulance indexAmb was completed;
\item
if ambulance indexAmb is currently at a station, then $t_{b}(\mbox{indexAmb})$ is the time when the ambulance arrived at the station, and if ambulance indexAmb is currently on its way to a station, then $t_{b}(\mbox{indexAmb})$ is the time that the ambulance is scheduled to arrive at the station, otherwise $t_{b}(\mbox{indexAmb})$ has a large value;
\item
if ambulance indexAmb is currently at a station, then $\ell_{b}(\mbox{indexAmb})$ is the location of the station, given by a (latitude,longitude) pair, and if ambulance indexAmb is currently on its way to a station, then $\ell_{b}(\mbox{indexAmb})$ is the location of that station;
\item
the $j$th location $\mbox{AmbulancesTrips}(\mbox{indexAmb})(j)$ visited by ambulance indexAmb during the simulation, given by a (latitude,longitude) pair (trip~$j$ of ambulance indexAmb starts at location $\mbox{AmbulancesTrips}(\mbox{indexAmb})(j)$ and ends at location \\ $\mbox{AmbulancesTrips}(\mbox{indexAmb})(j+1)$);
\item
the start time $\mbox{AmbulancesTimes}(\mbox{indexAmb})(j)$ of trip~$j$ of ambulance indexAmb;
\item
the type $\mbox{TripType}(\mbox{indexAmb})(j)$ of trip~$j$ of ambulance indexAmb, which can be one of the following types:
\begin{itemize}
\item
1: the ambulance is at an ambulance station (not to be cleaned);
\item
2: the ambulance is on its way to the scene of an emergency;
\item
3: the ambulance is on the scene of an emergency;
\item
4: the ambulance is on its way to a hospital;
\item
5: the ambulance is at a hospital, to transfer the patient(s);
\item
6: the ambulance is on its way to a cleaning station;
\item
7: the ambulance is being cleaned at a cleaning station;
\item
8: the ambulance is on its way to an ambulance station (not for a cleaning task).
\end{itemize}
Note that not all trip types involve movement of the ambulance.
The trip type numbers are in the same order as the tasks for an emergency, which starts with a trip of type 2 for an ambulance which was on a trip of type 1 or 8.
For example, if an ambulance transports a patient to a hospital and thereafter the ambulance has to be cleaned at a cleaning station, then the trip types for the corresponding emergency are, in chronological order, 2, 3, 4, 5, 6, 7, and 8.
\end{itemize}
The emergency variables and ambulance variables above are initialized with initial conditions when the simulation period starts.

As an example, Table~\ref{table_trajectory} shows the trip information described above for an emergency indexed with~$i$ and an ambulance indexed with indexAmb that serves that emergency.
\begin{itemize}
\item
Ambulance indexAmb arrives at station B at time $\mbox{AmbulancesTimes}(\mbox{indexAmb})(1)$ = 4:32. The index~1 in $\mbox{AmbulancesTimes}(\mbox{indexAmb})(1)$ means that it is the first simulated trip for ambulance indexAmb, and it also means that location 
$$\mbox{ambulancesTrip}(\mbox{indexAmb})(1)$$ is the (latitude,longitude) pair of station B.
\item
Ambulance indexAmb waits at station B until time $\mbox{AmbulancesTimes}(\mbox{indexAmb})(2)$ = 4:36, so that $\mbox{ambulancesTrip}(\mbox{indexAmb})(2)$ is the (latitude,longitude) pair of station B, and $\mbox{TripType}(\mbox{indexAmb})(1) = 1$.
\item
Ambulance indexAmb leaves station B at time $\mbox{AmbulancesTimes}(\mbox{indexAmb})(2)$ = 4:36 and arrives at the emergency location $\mbox{ambulancesTrip}(\mbox{indexAmb})(3) = \ell_{c}(i)$ at time $\mbox{AmbulancesTimes}(\mbox{indexAmb})(3)$ = 4:46, and $\mbox{TripType}(\mbox{indexAmb})(2) = 2$.
\item
Ambulance indexAmb stays at emergency location $\mbox{ambulancesTrip}(\mbox{indexAmb})(3) = \ell_{c}(i)$ until time $\mbox{AmbulancesTimes}(\mbox{indexAmb})(4)$ = 4:52, so that 
$$
\mbox{ambulancesTrip}(\mbox{indexAmb})(4) = \ell_{c}(i)$$ and $\mbox{TripType}(\mbox{indexAmb})(3) = 3.
$
\item
Ambulance indexAmb leaves emergency location $\mbox{ambulancesTrip}(\mbox{indexAmb})(3) = \ell_{c}(i)$ at time $\mbox{AmbulancesTimes}(\mbox{indexAmb})(4)$ = 4:52 and arrives at hospital H at time $\mbox{AmbulancesTimes}(\mbox{indexAmb})(5)$ = 5:06, so that $\mbox{ambulancesTrip}(\mbox{indexAmb})(5)$ is the (latitude,longitude) pair of hospital H and $\mbox{TripType}(\mbox{indexAmb})(4) = 4$.
\item
Ambulance indexAmb stays at hospital until time $\mbox{AmbulancesTimes}(\mbox{indexAmb})(6)$ = 5:25, so that $\mbox{ambulancesTrip}(\mbox{indexAmb})(6)$ is the (latitude,longitude) pair of hospital H and $\mbox{TripType}(\mbox{indexAmb})(5) = 5$.
\item
Ambulance indexAmb leaves hospital H at time $\mbox{AmbulancesTimes}(\mbox{indexAmb})(6)$ = 5:25 and arrives at station B at time time $\mbox{AmbulancesTimes}(\mbox{indexAmb})(7)$ = 5:45, so that $\mbox{ambulancesTrip}(\mbox{indexAmb})(7)$ is the (latitude,longitude) pair of station B and $\mbox{TripType}(\mbox{indexAmb})(6) = 6$.
\end{itemize}

\begin{table}[hbtp]
\centering
\begin{tabular}{cccccccccccccc}
\hline
TripType  & & 1 && 2 && 3 && 4 && 5 && 6 & \\ 
\hline
AmbulancesTrips & B && B & &$\ell_{c}(i)$ && $\ell_{c}(i)$ && H && H && B  \\ 
\hline
AmbulancesTimes    & 4:32 && 4:36 && 4:46 && 4:52 && 5:06 && 5:25 && 5:45\\ 
\hline
\end{tabular}
\caption{Example of the trips of an ambulance for an emergency from station B to the emergency location $\ell_{c}(i)$, then to hospital H, and back to station B.}
\label{table_trajectory}
\end{table}

\begin{figure}
\centering
\begin{tabular}{c}
\includegraphics[scale=1.07]{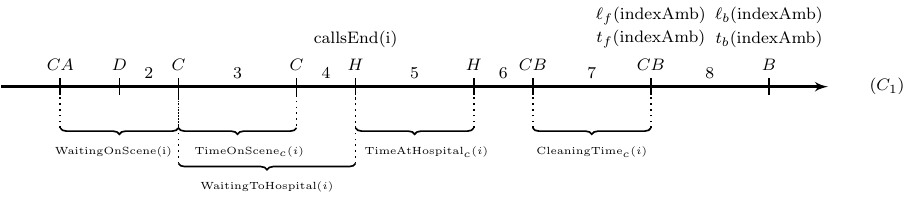}\\
\includegraphics[scale=1.12]{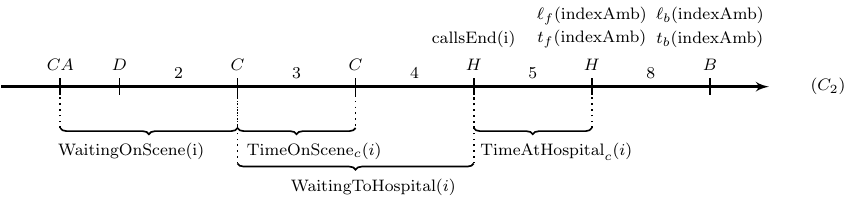}\\
\includegraphics[scale=1.1]{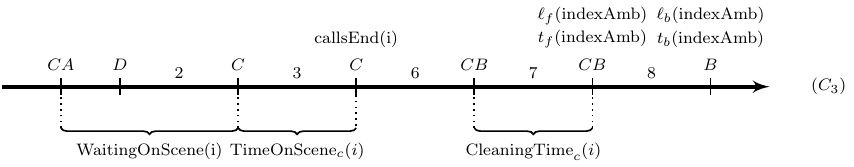}\\
\includegraphics[scale=1.18]{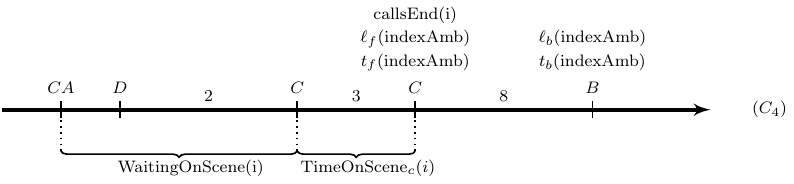}
\end{tabular}
\caption{Sequences of trips for the classes of service $C_{1}$, $C_{2}$, $C_{3}$, $C_{4}$.
The letters above the time axis represent locations for the corresponding time instants.
Specifically, CA: call arrival, D: departure of the ambulance to the emergency scene, C: emergency scene, H: hospital, CB: cleaning station, B: staging station.
Also, the trip type numbers 2, 3, 4, 5, 6, 7, and 8 are shown above the time axis for all classes of service.}
\label{figgroupcalls}
\end{figure}

There are four classes of service, labeled $C_{1}$, $C_{2}$, $C_{3}$, $C_{4}$, depending on whether the ambulance transports the patient(s) to a hospital or not and whether the ambulance travels to a cleaning station after service or not.
Figure~\ref{figgroupcalls} shows the sequences of trip types for the four classes of service.

For any two locations $A, B$, given by (latitude,longitude) pairs, let
\begin{equation}
\label{tratime}
{\mbox{travelTime}}(A, B, t_{0})
\end{equation}
denote the travel time from $A$ to $B$ of an ambulance trip which starts at $t_{0}$.
For any time instant $t \in [t_{0}, t_{0} + {\mbox{travelTime}}(A, B, t_{0})]$, let
\begin{equation}
\label{posbetweenordest}
{\mbox{positionBetweenOriginDestination}}(A, B, t_{0}, t,v)
\end{equation}
denote the position, given by a (latitude,longitude) pair, at time $t$, of an ambulance that starts a trip from $A$ to $B$ at time instant $t_{0}$ and travels at constant speed~$v$ along the great circle path from $A$ to $B$.

Next we describe how the simulation tracks ambulance trips.
Consider an emergency~$i$ served by ambulance indexAmb, which is busy with trip~$j$ before it serves emergency~$i$.
The following three cases are distinguished: (A) $t_{b}(\mbox{indexAmb}) \leq t_{c}(i)$, (B) $t_{f}(\mbox{indexAmb}) \leq t_{c}(i) < t_{b}(\mbox{indexAmb})$, and (C) $t_{c}(i) < t_{f}(\mbox{indexAmb})$:
\begin{enumerate}
\item[(A)]
$t_{b}(\mbox{indexAmb}) \leq t_{c}(i)$, i.e., ambulance~$\mbox{indexAmb}$ is at a station when call~$i$ arrives.
Then $\mbox{TripType}(\mbox{indexAmb})(j) = 1$, $\mbox{AmbulancesTimes}(\mbox{indexAmb})(j) = t_{b}(\mbox{indexAmb})$,
$$
{\small{
\mbox{AmbulancesTrips}(\mbox{indexAmb})(j) = \mbox{AmbulancesTrips}\\(\mbox{indexAmb})(j+1) = \ell_{b}(\mbox{indexAmb}),
}}
$$
and $\mbox{AmbulancesTimes}(\mbox{indexAmb})(j+1) = t_{c}(i)$.
Also, $\mbox{TripType}(\mbox{indexAmb})(j+1) = 2$, $\mbox{AmbulancesTrips}(\mbox{indexAmb})(j+2) = \ell_{c}(i)$, and $\mbox{AmbulancesTimes}(\mbox{indexAmb})(j+2) = t_{c}(i) + \mbox{travelTime}(\ell_{b}(\mbox{indexAmb}),\ell_{c}(i),t_{c}(i))$.
In this case, the response time is the time for the ambulance to go from station $\ell_{b}(\mbox{indexAmb})$ to the emergency location $\ell_{c}(i)$, starting the trip at time $t_{c}(i)$.
Thus, the response time is $\mbox{WaitingOnScene}(i) = \mbox{travelTime}(\ell_{b}(\mbox{indexAmb}),\ell_{c}(i),t_{c}(i))$.
\item[(B)]
$t_{f}(\mbox{indexAmb}) \leq t_{c}(i) < t_{b}(\mbox{indexAmb})$, i.e., the ambulance has completed its previous service and is currently on its way to a station.
Then the current position~$P$ of ambulance indexAmb is given by 
\[
\begin{array}{lcl}
P & =& \mbox{positionBetweenOriginDestination}\big(\ell_{f}(\mbox{indexAmb}), \ell_{b}(\mbox{indexAmb}),\\
&& \hspace*{6.4cm}t_{f}(\mbox{indexAmb}),t_{c}(i),v\big).
\end{array}
\]
Then $\mbox{TripType}(\mbox{indexAmb})(j) = 8$, $\mbox{AmbulancesTrips}(\mbox{indexAmb})(j) = \ell_{f}(\mbox{indexAmb})$, $\mbox{AmbulancesTrips}(\mbox{indexAmb})(j+1) = P$, 
$$\mbox{AmbulancesTimes}(\mbox{indexAmb})(j) = t_{f}(\mbox{indexAmb}),
$$
and $\mbox{AmbulancesTimes}(\mbox{indexAmb})(j+1) = t_{c}(i)$.
Also, $\mbox{TripType}(\mbox{indexAmb})(j+1) = 2$, $\mbox{AmbulancesTrips}(\mbox{indexAmb})(j+2) = \ell_{c}(i)$, and $\mbox{AmbulancesTimes}(\mbox{indexAmb})(j+2) = t_{c}(i) + \mbox{travelTime}(P,\ell_{c}(i),t_{c}(i))$.
The response time is then the time for ambulance indexAmb to travel from location $P$ at time $t_{c}(i)$ to location $\ell_{c}(i)$.
Thus, the response time is $\mbox{WaitingOnScene}(i) = \mbox{travelTime}(P, \ell_{c}(i), t_{c}(i))$.
\item[(C)]
$t_{c}(i) < t_{f}(\mbox{indexAmb})$, i.e., ambulance~$\mbox{indexAmb}$ is serving another emergency when call~$i$ arrives.
This includes the case in which emergency~$i$ was put in queue when it arrived, and is later served by ambulance indexAmb.
Then the scheduled time until the ambulance completes its previous service is equal to $t_{f}(\mbox{indexAmb}) - t_{c}(i)$.
Thereafter, the ambulance travels from location $\ell_{f}(\mbox{indexAmb})$ to the call location.
Then 
$
\mbox{AmbulancesTrips}(\mbox{indexAmb})(j+1) = \ell_{f}(\mbox{indexAmb}),
$
$\mbox{TripType}(\mbox{indexAmb})(j) \in \{2,3,4,5,6,7\}$,  $\mbox{AmbulancesTimes}(\mbox{indexAmb})(j+1) = t_{f}(\mbox{indexAmb})$.

Also, $\mbox{TripType}(\mbox{indexAmb})(j+1) = 2$, $\mbox{AmbulancesTrips}(\mbox{indexAmb})(j+2) = \ell_{c}(i)$, and 
$$
\begin{array}{l}
\mbox{AmbulancesTimes}(\mbox{indexAmb})(j+2) = t_{f}(\mbox{indexAmb})\\
+ \mbox{travelTime}(\ell_{f}(\mbox{indexAmb}),\ell_{c}(i),t_{f}(\mbox{indexAmb})).
\end{array}
$$
Thus, the response time is
\[
\begin{array}{lcl}
\mbox{WaitingOnScene}(i) & = & t_{f}(\mbox{indexAmb}) - t_{c}(i)\\ 
&&+ \mbox{travelTime}\big(\ell_{f}(\mbox{indexAmb}),\ell_{c}(i),t_{f}(\mbox{indexAmb})\big).
\end{array}
\]
\end{enumerate}
In all three cases, after ambulance indexAmb arrives on the scene of emergency~$i$, it spends amount of time $\mbox{TimeOnScene}(i)$ at the scene of the emergency.

Thus, $\mbox{TripType}(\mbox{indexAmb})(j+2) = 3$, $\mbox{AmbulancesTrips}(\mbox{indexAmb})(j+3) = \ell_{c}(i)$, and 
$$
\begin{array}{lcl}
\mbox{AmbulancesTimes}(\mbox{indexAmb})(j+3) &=& \mbox{AmbulancesTimes}(\mbox{indexAmb})(j+2)\\
&&+ \mbox{TimeOnScene}(i).
\end{array}
$$

We show the computations only for class of service~$C_{1}$ in which the ambulance goes to a hospital and after that to a cleaning station.
Computations for the other classes are easy modifications of computations for this class.

After ambulance indexAmb leaves the scene of emergency~$i$, it travels to a hospital at location $\ell_{h}(i)$.
Thus, $\mbox{TripType}(\mbox{indexAmb})(j+3) = 4$, $\mbox{AmbulancesTrips}(\mbox{indexAmb})(j+4) = \ell_{h}(i)$, and $\mbox{AmbulancesTimes}(\mbox{indexAmb})(j+4) = \mbox{AmbulancesTimes}(\mbox{indexAmb})(j+3) + \mbox{travelTime}(\ell_{c}(i),\ell_{h}(i),\mbox{AmbulancesTimes}(\mbox{indexAmb})(j+3))$.
Also, $\mbox{WaitingToHospital}(i) = \mbox{TimeOnScene}(i) + \mbox{travelTime}(\ell_{c}(i),\ell_{h}(i),\mbox{AmbulancesTimes}(\mbox{indexAmb})(j+3))$.

The ambulance remains at the hospital to transfer the patient(s) to the emergency department for an amount of time $\mbox{TimeAtHospital}(i)$.
Thus, $\mbox{TripType}(\mbox{indexAmb})(j+4) = 5$, $\mbox{AmbulancesTrips}(\mbox{indexAmb})(j+5) = \ell_{h}(i)$, and $\mbox{AmbulancesTimes}(\mbox{indexAmb})(j+5) = \mbox{AmbulancesTimes}(\mbox{indexAmb})(j+4) + \mbox{TimeAtHospital}(i)$.

Next, ambulance indexAmb travels from the hospital to a cleaning station at location $\ell_{cb}(i)$ for a thorough cleaning.
Thus, $\mbox{TripType}(\mbox{indexAmb})(j+5) = 6$, $\mbox{AmbulancesTrips}\\(\mbox{indexAmb})(j+6) = \ell_{cb}(i)$, and $\mbox{AmbulancesTimes}(\mbox{indexAmb})(j+6) = \mbox{AmbulancesTimes}\\(\mbox{indexAmb})(j+5) + \mbox{travelTime}(\ell_{h}(i),\ell_{cb}(i),\mbox{AmbulancesTimes}(\mbox{indexAmb})(j+5))$.

The ambulance remains at the cleaning station for an amount of time $\mbox{CleaningTime}(i)$ until it has been cleaned.

Thus, $\mbox{TripType}(\mbox{indexAmb})(j+6) = 7$, $\mbox{AmbulancesTrips}(\mbox{indexAmb})(j+7) = \ell_{cb}(i)$, and 
$$
\begin{array}{lcl}
\mbox{AmbulancesTimes}(\mbox{indexAmb})(j+7) &= &\mbox{AmbulancesTimes}(\mbox{indexAmb})(j+6)\\
&&+ \mbox{CleaningTime}(i).
\end{array}
$$
At this time, the service of emergency~$i$ is completed, and thus state variable $t_{f}(\mbox{indexAmb})$ is updated to $\mbox{AmbulancesTimes}(\mbox{indexAmb})(j+7)$ and state variable $\ell_{f}(\mbox{indexAmb})$ is updated to $\ell_{cb}(i)$.

Next, ambulance indexAmb travels from the cleaning station to the ambulance station at $B(i)$ to wait for its next assignment.
State variable $t_{b}(\mbox{indexAmb})$ is updated to 
$$
\begin{array}{l}
\mbox{AmbulancesTimes}(\mbox{indexAmb})(j+7)\\
+ \mbox{travelTime}(\ell_{cb}(i),B(i),\mbox{AmbulancesTimes}(\mbox{indexAmb})(j+7))
\end{array}
$$ and state variable $\ell_{b}(\mbox{indexAmb})$ is updated to $B(i)$.
If the ambulance travels all the way to $B(i)$ before its next assignment, then $\mbox{TripType}(\mbox{indexAmb})(j+7) = 8$, $\mbox{AmbulancesTrips}\\(\mbox{indexAmb})(j+8) = B(i)$, and $\mbox{AmbulancesTimes}(\mbox{indexAmb})(j+8) = \mbox{AmbulancesTimes}\\(\mbox{indexAmb})(j+7) + \mbox{travelTime}(\ell_{cb}(i),B(i),\mbox{AmbulancesTimes}(\mbox{indexAmb})(j+7))$.
The ambulance can be dispatched to an arriving emergency, say emergency~$i'$, while on its way to $B(i)$.
In such a case, $\mbox{TripType}(\mbox{indexAmb})(j+7) = 8$, $\mbox{AmbulancesTrips}(\mbox{indexAmb})(j+8) = P$, and $\mbox{AmbulancesTimes}(\mbox{indexAmb})(j+8)$ is determined as in case~(B) above.
A cleaning station can also serve as a staging station, so it is possible that $B(i) = \ell_{cb}(i)$.
In such a case, $\mbox{TripType}(\mbox{indexAmb})(j+7) = 1$, $\mbox{AmbulancesTrips}(\mbox{indexAmb})(j+8) = B(i) = \ell_{cb}(i)$, and $\mbox{AmbulancesTimes}(\mbox{indexAmb})(j+8)$ is determined as in case~(A) above.
As discussed before, if there are emergencies waiting in queue, then it is also possible that the ambulance is assigned to another emergency, say emergency~$i'$, after cleaning.
In such a case, $\mbox{TripType}(\mbox{indexAmb})(j+7) = 2$, $\mbox{AmbulancesTrips}(\mbox{indexAmb})(j+8) = \ell_{c}(i')$, and $\mbox{AmbulancesTimes}(\mbox{indexAmb})(j+8)$ is determined as in case~(C) above.

\subsection{Performance metrics}

Two considerations are represented in the performance metrics:
\begin{enumerate}
\item
Response times are important in many emergencies.
The impact of response time on patient well-being is different for different types of emergencies.
Also, for some emergencies the time until an ambulance arrives on the scene is more important, whereas for other emergencies the time until the patient is treated at a hospital is more important.
\item
Different ambulances and different crews have different capabilities, and the impact of ambulance and crew capabilities on patient well-being is different for different types of emergencies.
\end{enumerate}
Therefore, the performance metric cost\_allocation\_ambulance consists of two terms: a term that penalizes response time depending on the type of emergency, and a term that penalizes the mismatch of ambulance and crew capabilities with the needs of the type of emergency.
More specifically, of ``cost'' of allocating an ambulance of type~$a$ to an emergency of type $c$ with response time~$t$ is given by
\begin{equation}
\label{costalloc}
\mbox{cost\_allocation\_ambulance}(a,c,t) \ \ = \ \ \mbox{{\tt{penalization}}}(t,c) + M_{ac}
\end{equation}
In \eqref{costalloc}, 
\begin{itemize}
\item
{\tt{penalization}}(t,c) is the penalty if an emergency of type~$c$ is served with response time~$t$.
Simulation results currently displayed on the webpage uses
\begin{equation}
\label{penbtct}
\mbox{{\tt{penalization}}}(t,c) \ \ = \ \ \theta_{c} t
\end{equation}
where $\theta_{c}$ is a coefficient that depends on the emergency type~$c$.
\item
$M_{ac}$ is the cost of assigning an ambulance of type~$a$ to an emergency of type $c$. 
\end{itemize}


\subsection{Ambulance selection and reassignment heuristics}
\label{sec:heur}

In this section, we provide a brief description of the ambulance selection and ambulance reassignment heuristics available through the webpage \url{http://samu.vincentguigues.com/} (the paper
\cite{ourheuristics23a} contains more details and performance comparisons for these heuristics).
These heuristics are the Closest Available (CA) heuristic, Best Myopic (BM) heuristic, GHP1 (Greedy Heuristic with Priorities 1), and GHP2 (Greedy Heuristic with Priorities 2).

For all heuristics, the state vector will store, at any time $t$:
\begin{enumerate}
\item[1)]
for each ambulance $j$, a location $\ell_{f}(j)$ and a time $t_{f}(j)$ with the following meanings.
Two situations can happen:
1.1) an ambulance $j$ is in service at $t$ or 1.2) it is available at $t$.
\par 1.1) If the ambulance is in service at $t$ then $t_{f}(j)$ is the next instant the ambulance will be available (free) again for dispatch and $\ell_{f}(j)$ will be its location at that instant.
\par 1.2) If the ambulance is available at $t$ then $t_{f}(j)$ is the last (past) time instant the ambulance became available, i.e., the time it completed its last service and  $\ell_{f}(j)$ was the corresponding (past) location of the ambulance when this service was completed.
\item[2)]
For each ambulance $j$ a location $\ell_{b}(j)$ and a time $t_{b}(j)$ with the following meanings.
We again have two possibilities.
Either the ambulance is at a base at $t$ and, in this case, $\ell_{b}(j)$ is the location of that base and $t_{b}(j)$ is the last time instant it arrived at this base (therefore $t_{b}(j) \leq t$).
If the ambulance is not at a base, then $\ell_{b}(j)$ is the location of the next base it will go and $t_{b}(j)$ is the instant the ambulance will arrive at that base (we therefore have $t_{b}(j) > t$).
\end{enumerate}

\paragraph{Closest Available (CA) ambulance heuristic.}
One of the most popular heuristics for ambulance selection in the literature, called the Closest Available ambulance heuristic, works as follows.
When a call arrives, if there is no ambulance available, then the emergency is put in a queue of calls (many papers assume that in such a case the emergency is handled by an ``outside'' agency, and the emergency is removed from the system, so that in these papers there is no queue).
Otherwise, among the available ambulances, the closest (in time) is sent to the emergency.
When an ambulance becomes available, if there are calls in queue, then we send the ambulance to the oldest call in queue (as mentioned above, in many papers there is no queue, so that this case does not apply).
Otherwise, the ambulance is sent to the closest ambulance station or to its home base if the home base rule is used.

\paragraph{Best Myopic (BM) heuristic.}
When a call arrives, the BM heuristic computes the allocation cost given by \eqref{costalloc} for every ambulance given the current state of all ambulances, and the ambulance with the least allocation cost is selected to respond to the emergency.
If there is a tie among the ambulances achieving the least allocation cost, then the heuristic selects the least advanced ambulance (for example if there is a BLS and an ALS ambulance achieving the smallest allocation cost, then the heuristic sends the BLS ambulance).
When an ambulance becomes available, if there are calls in queue, then the ambulance is sent to the oldest call in queue.
Otherwise, the ambulance is sent to the closest ambulance station or to its home base if the home base rule is used.

\if{

\paragraph{NM heuristic.}
The NM heuristic assumes that we know, potentially, all the calls that will arrive in a given future time window, but we do not send ambulances to calls before these calls actually arrive.
Under uncertainty in the time and location of calls, it can be used in a rolling horizon approach to solve second stage scenarios as in Section 3 of \cite{ourheuristics23a}.
For each call \(i\) for which no ambulance has been sent, NM computes the set \(\mathcal{S}(i)\) of best ambulances which are the ambulances that achieve the minimal allocation cost, given previous ambulance allocations, where the allocation cost is given by \eqref{costalloc}.
If any ambulances in \(\mathcal{S}(i)\) are available, then NM dispatches a least advanced ambulance (as basic as possible) from \(\mathcal{S}(i)\).
Otherwise, we compute for each ambulance \(j \in \mathcal{S}(i)\), the set \(\mathcal{T}(j)\) of calls that arrive not later than \(t_{f}(j)\) and such that \(j\) is one of the best ambulances for that call, in terms of allocation cost.
An ambulance \(j \in \mathcal{S}(i)\) is termed {\em{good}} for call \(i\) if, for every call \(k \in T(j)\), the cost of allocating \(j\) to \(i\) is less than or equal to the cost of allocating \(j\) to \(k\).
If the ambulance \(j\) is good for call \(i\), we dispatch \(j\) to \(i\) after it finishes its current service.
Otherwise, we dispatch \(j\) to the call \(k \in T(j)\) that achieves the minimal allocation cost.
When an ambulance after service is not immediately sent to an emergency, it is sent to its home base (if the home base rule is used) or to the closest station.

}\fi

\paragraph{GHP1 heuristic.}
In GHP1, every time a call arrives, it is put in the queue of calls, possibly for immediate dispatch, see below.
Given the current time $t$ (this is a time at which we need to make a decision: when a call arrives or when an ambulance finishes service), we sort the queue of calls by decreasing order of the penalized waiting time.
Specifically, for call $i$ of type $\mbox{Type}_{c}(i)$ that arrived at time $t_{c}(i)$, the penalized waiting time is {\tt{penalization}}$(t - t_{c}(i),\mbox{Type}_{c}(i))$.
Considering the calls in decreasing order of the penalized waiting time, we do for every call the following: we compute the ambulances that achieve the smallest allocation cost for that call.
If any of them are available, we send to that call, among these ambulances, an ambulance as basic as possible.
Otherwise the call remains in the queue of calls, and we proceed to the next call in queue.
When an ambulance is sent to a station, it goes either to the closest station or to its home base.

\paragraph{GHP2 heuristic.}
When a new call arrives, we put this call in the queue of calls.
We then go through all calls in queue as follows.
We compute for every call~$k$ in queue the minimal allocation cost minAlloc(k) given by \eqref{costalloc} and the calls that achieve the maximal value of minAlloc(k) among all calls k.
If all ambulances of all these calls are not available, then we select one of these calls arbitrarily.
An ambulance will be dispatched for that call at a later time (this call is put in the new queue of calls).
Otherwise, let currentCall be a call for which there is at least one available ambulance among the best ambulances (with minimal allocation cost) for that call.
We send to call currentCall, among these ambulances, an ambulance as basic as possible.
We update state vectors, ambulance rides, waiting times and allocation costs (counted from the current time $t$ and given previously scheduled ambulance rides).
We then go to the next call in queue and repeat the procedure described previously.
In GHP2 again, when an ambulance is sent to a station, it goes either to the closest station or to its home base.

\subsection{Upload of new EMS data}

\begin{figure}
\centering
\begin{tabular}{c}
\includegraphics[scale=0.2]{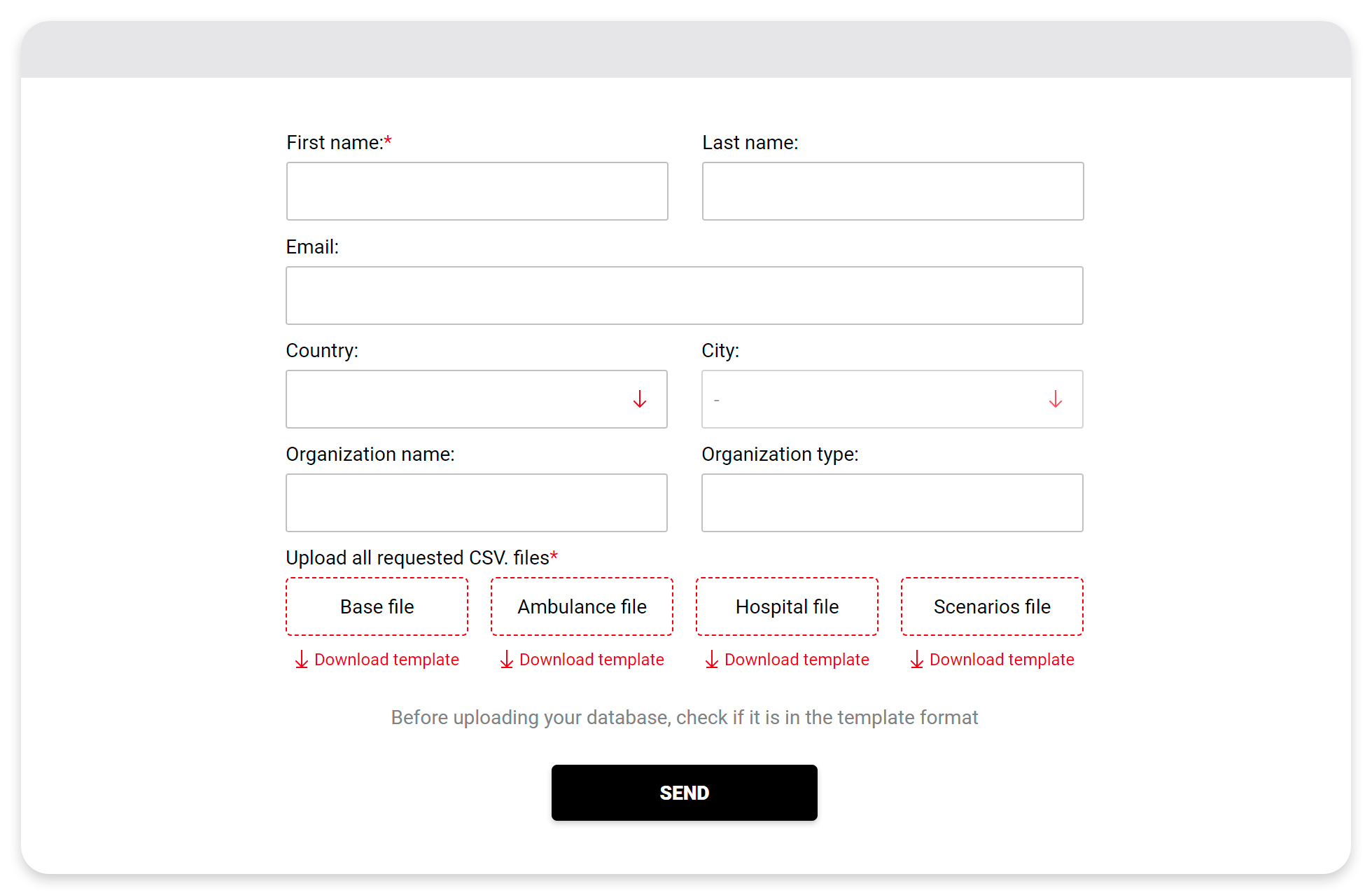}
\end{tabular}
\caption{Form to send new data of EHS.}\label{fig9}
\end{figure}

New EMS system data can be uploaded clicking the link
{\em{SAMU - Response Times}}
then the link 
{\em{Distribution of Response Times}}
and clicking the
{\em{Upload data}} button
which opens the form shown in Figure~\ref{fig9}. 
Four types of data can be uploaded: (1)~the locations of ambulance stations, (2)~the list of ambulances and their home bases (if applicable), (3)~the locations of hospitals, and (4)~scenarios of emergency calls.
Templates for these files are provided.

\subsection{Visualization of ambulance movements}
\label{sec:discamb}

The arrival of emergencies, and the selection of movements of ambulances under a chosen policy can be simulated and visualized on the website.

The link {\em{Trajectories simulation}} available clicking the link 
{\em{SAMU-Response Times}}, provides the visualization on a map of ambulance trajectories simulated with our allocation policies
on a set of scenarios.

\begin{figure}
\centering
\begin{tabular}{c}
\includegraphics[scale=0.19]{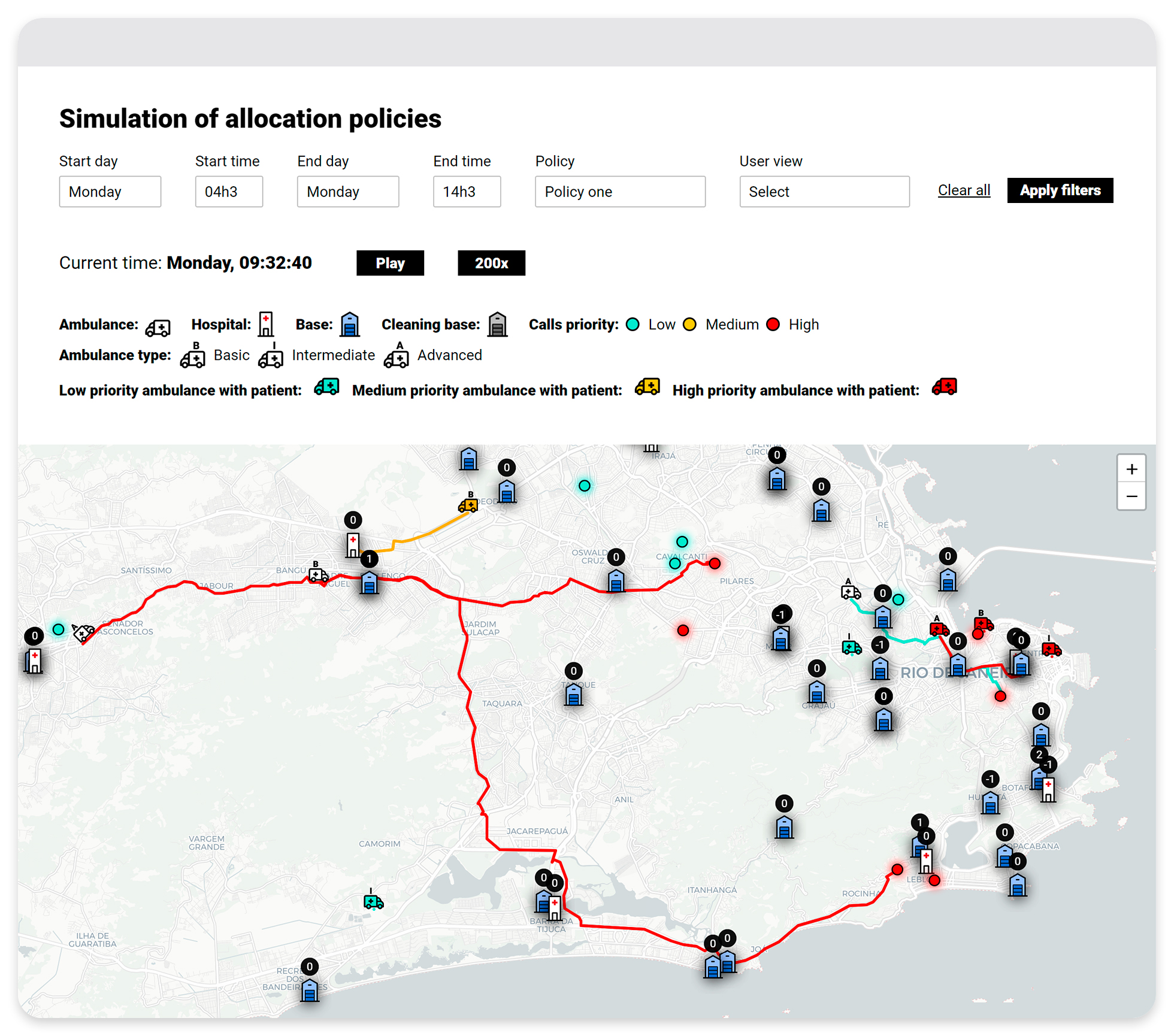}
\end{tabular}
\caption{Visualization of ambulance trajectories in admin view.}\label{fig13}
\end{figure}

\begin{figure}
\centering
\begin{tabular}{c}
\includegraphics[scale=0.16]{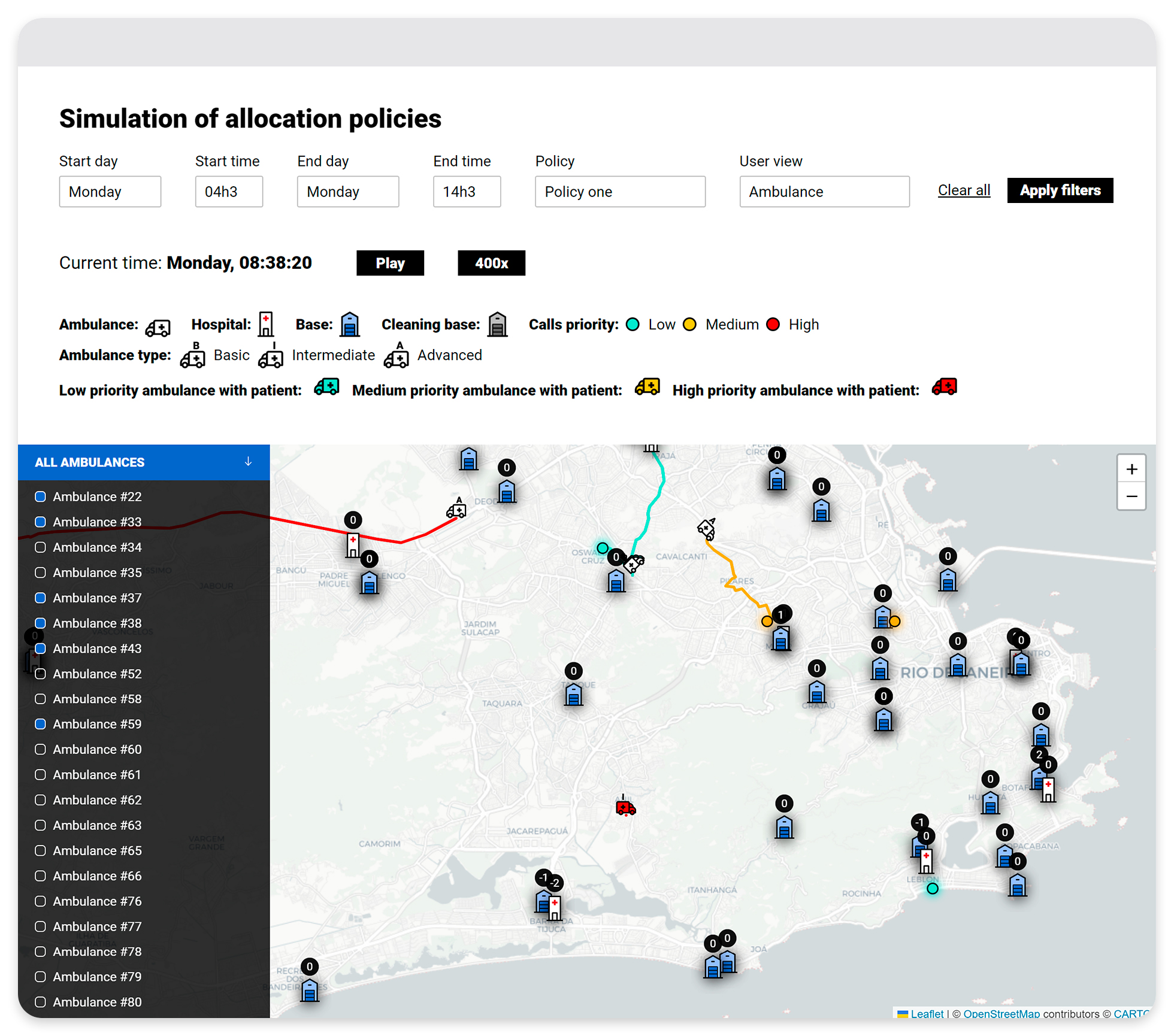}
\end{tabular}
\caption{Visualization of ambulance trajectories in ambulance view.}\label{fig14}
\end{figure}

\begin{figure}
\centering
\begin{tabular}{c}
\includegraphics[scale=0.16]{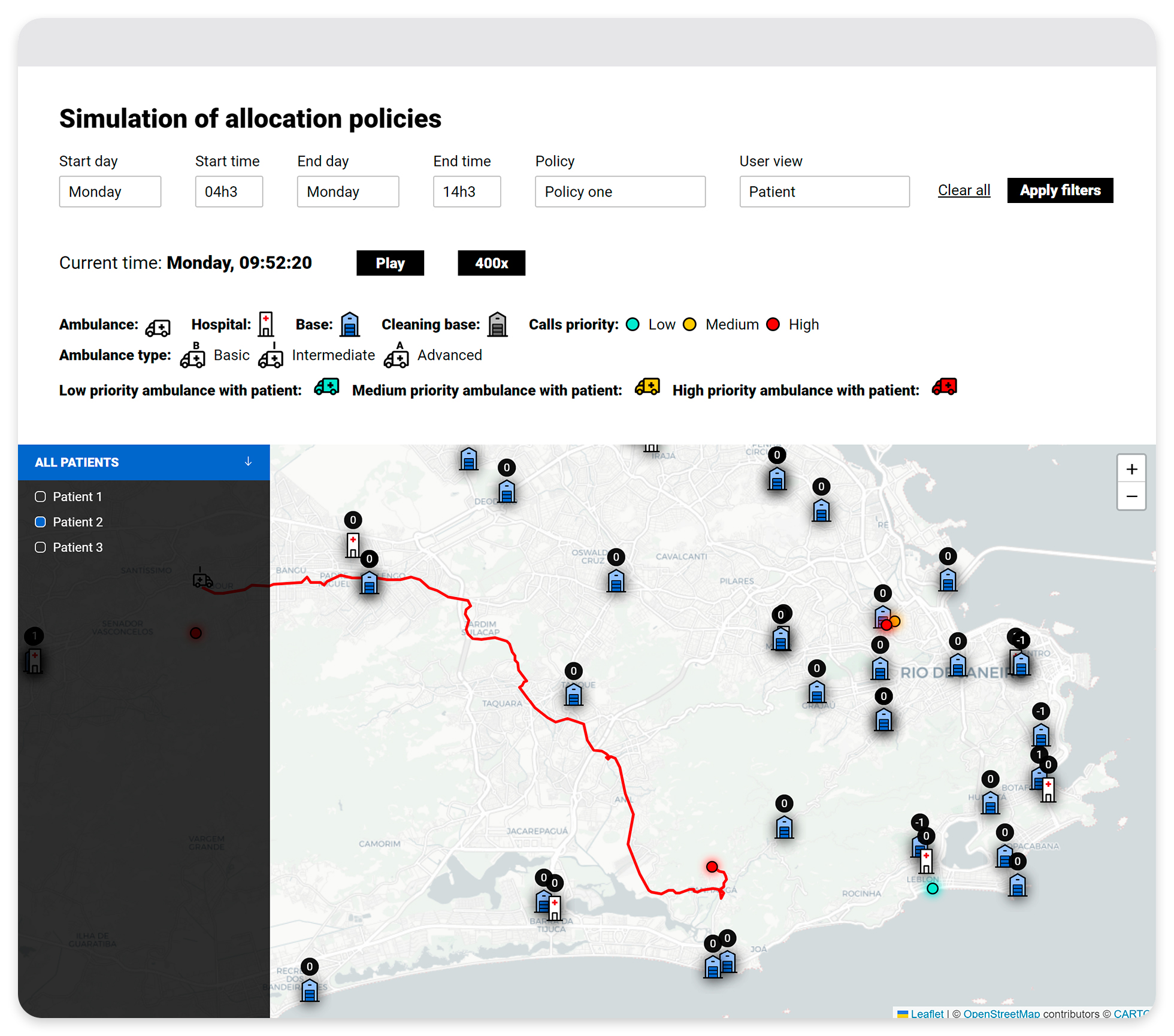}
\end{tabular}
\caption{Visualization of ambulance trajectories in patient view.}\label{fig15}
\end{figure}

A set of filters lets the user choose the time window of the simulation (start day, end day, start time, end time), as well as one of the provided policies for ambulance selection and reassignment (the currently provided policies are described in Section~\ref{sec:heur}).
The simulation randomly generates emergency calls from the distribution of calls for the chosen time period.
The simulation of the resulting ambulance operations is described in Section~\ref{sec:ambfleet}.
The website provides three visualization modes: (1)~admin view, that shows all ambulances and emergencies; (2)~ambulance view, in which the user can select a subset of ambulances to track; and (3)~patient view, in which the user can select a subset of emergencies to track.
An example of the admin view is shown in Figure~\ref{fig13}, an example of the ambulance view is shown in Figure~\ref{fig14}, and an example of the patient view is shown in Figure~\ref{fig15}.

The user can also choose the time acceleration factor: $\times 2$, $\times 10$, $\times 50$, $\times 100$, $\times 200$, or $\times 400$.
For example, the acceleration factor $\times 100$ runs a simulation for a time window of $120$~minutes in $120/100$~minutes or $72$~seconds of user time.
Ambulance trajectories are updated every $t_{\mathrm{step}}$~seconds of simulated time using the computations described in Section~\ref{sec:ambulance trajectories}.
Thus, with acceleration factor $\times 100$, ambulance trajectories are updated every $t_{\mathrm{step}}/100$~seconds of user time.
A legend is given on the map, which shows $12$~types of information:
\begin{itemize}
\item
hospitals; 
\item
ambulance stations;
\item
low priority emergencies;
\item
intermediate priority emergencies;
\item
high priority emergencies;
\item
basic life support ambulances with patient(s);
\item
intermediate life support ambulances with patient(s);
\item
advanced life support ambulances with patient(s);
\item
basic life support ambulances without patient(s);
\item
intermediate life support ambulances without patient(s);
\item
advanced life support ambulances without patient(s);
\item
lines showing future trajectories of ambulances while on service.
These lines follow the streets and go from the current ambulance location until
the next stop (such as emergency scene, hospital, or ambulance station).
\end{itemize}

\subsection{Computation of ambulance trajectories}
\label{sec:ambulance trajectories}

In this section we describe how the ambulance position is computed every $t_{\mathrm{step}}$ simulated seconds (for example, if $t_{\mathrm{step}} = 5$, then we compute and visualize on a map the position of each ambulance every $5$~seconds of simulated time).
Given the graph of the streets of the considered region, for every trip in arrays AmbulancesTimes, AmbulancesTrips, and TripType, say from origin O to destination D, we compute the shortest path from O to D, travelling along the streets of the city.
This path is given by a sequence of nodes where the first node corresponds to origin O and the last node corresponds to destination D.
Using these shortest paths, and arrays AmbulancesTrips, AmbulancesTimes, and TripType, we compute arrays Trips, Times, and Types, where Trips contains the nodes of the street graph visited by the ambulance along the shortest path, Times contains the time instants when the ambulances arrive at these nodes, and Types contains the types of all trips between two consecutive nodes.
Finally, Algorithm \ref{algo_discretization} is used to compute arrays discretized\_rides, discretized\_ride\_times, and discretized\_ride\_types, where discretized\_rides contains the locations of each ambulance every $t_{\mathrm{step}}$ simulated seconds, discretized\_ride\_times contains the time instants when the ambulances arrive at these locations, and discretized\_ride\_types contains the types of all trips between two consecutive locations.
In this algorithm, for each ambulance indexAmb and every time instant~$t$ that is a multiple of $t_{\mathrm{step}}$ within the time window of the simulation, we find the index~$i$ such that $\mbox{Times}(\mbox{indexAmb})(i) \le t < \mbox{Times}(\mbox{indexAmb})(i+1)$, that is, time instant~$t$ is between time instants $t_{\mathrm{prev}} = \mbox{Times}(\mbox{indexAmb})(i)$ (when ambulance indexAmb is at $P_{1} = \mbox{Trips}(\mbox{indexAmb})(i)$) and $t_{\mathrm{next}} = \mbox{Times}(\mbox{indexAmb})(i+1)$ (when ambulance indexAmb is at $P_{2} = \mbox{Trips}(\mbox{indexAmb})(i+1)$).
We then find the position~$P$ of the ambulance assuming that it goes from $P_{1}$ to $P_{2}$ along the great circle path at constant speed~$v$, that is, $P = \mbox{positionBetweenOriginDestination}(P_{1},P_{2},t_{\mathrm{prev}},t)$.
Next we explain how to compute the latitude and longitude of such point $P$ given $P_{1} = (\ell_{1},L_{1})$ and $P_{2} = (\ell_{2},L_{2})$ (latitude,longitude).

\begin{figure}
\centering
\begin{tabular}{c}
\includegraphics[scale=1.3]{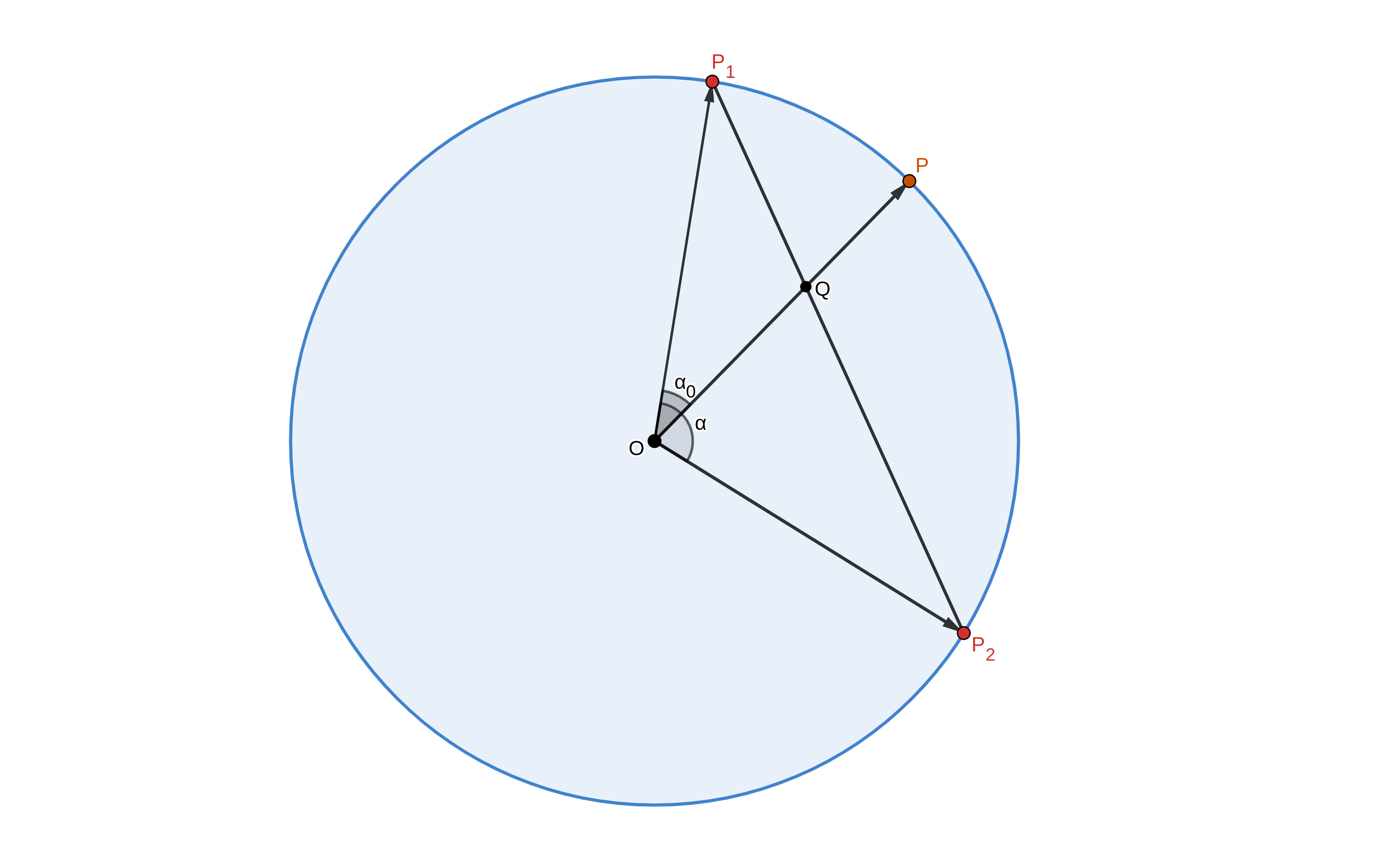}\\
\includegraphics[scale=0.38]{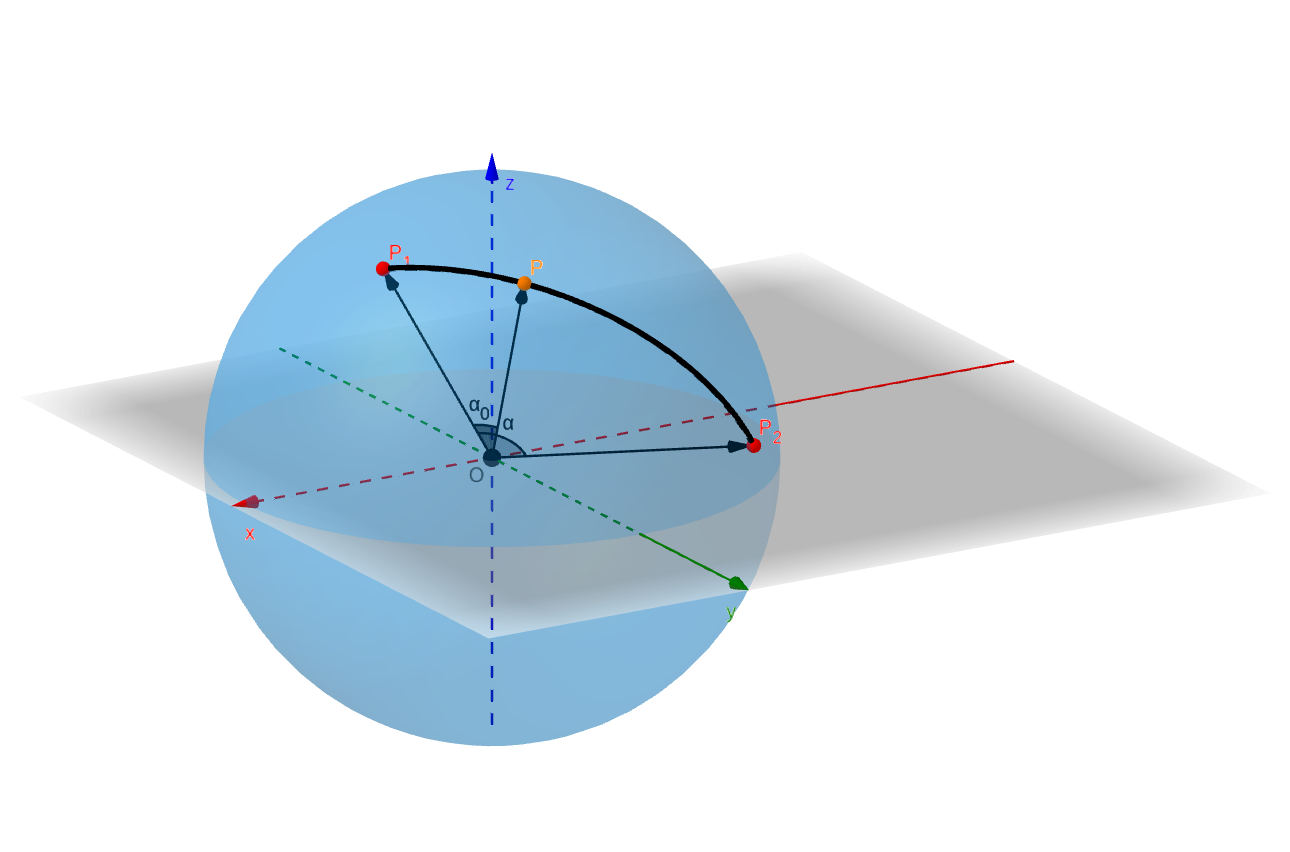}
\end{tabular}
\caption{Bottom plot: geodesic from $P_{1}$ to $P_{2}$ with $\alpha=\angle P_{1}OP_{2}$ and point $P$ on this geodesic with $\alpha_0 = \angle P_{1} O P$. On the 2-D top plot, we also represented the intersection $Q$ of $[O,P]$ and $[P_{1},P_{2}]$\label{figuregeodesic}
}
\end{figure}

Recall that latitudes are between $-90^{\circ}$ and $90^{\circ}$ while longitudes are between $-180^{\circ}$ and $180^{\circ}$.
The Cartesian coordinates of $P_{1}$ and $P_{2}$ are given by
\[
P_{i} \ \ = \ \ \left(
\begin{array}{c}
x_{i}\\
y_{i}\\
z_{i}
\end{array}
\right)
\ \ = \ \
\left(
\begin{array}{c}
R \cos \left( \pi \ell_{i} /180 \right) \cos\left( \pi L_{i} / 180 \right)\\
R \cos \left( \pi \ell_{i} / 180 \right) \sin \left( \pi L_{i} / 180 \right)\\
R \sin \left( \pi \ell_{i} / 180 \right) 
\end{array}
\right)
\]
for $i=1,2$, where $R = 6371$km is the average radius of the earth.
The time for the ambulance to go from $P_{1}$ to $P_{2}$ at constant speed $v$ is $\Delta = 2R \arcsin(\|P_{2} - P_{1}\| / 2R) / v$.
Therefore, if $t \geq t_{\mathrm{prev}} + \Delta$, then the ambulance arrives at destination $P_{2}$ no later than $t$.
If $t < t_{\mathrm{prev}} + \Delta$, then to compute the latitude and longitude of $P$, we refer to Figure~\ref{figuregeodesic}, where we have represented $P_{1}$, $P_{2}$, $P$, the intersection $Q$ between $[P_{1},P_{2}]$ and $[O,P]$ (O is the center of the earth), angle $\alpha = \angle P_{1} O P_{2} = 2 \arcsin(\|P_{2} - P_{1}\| / 2R)$, and $\alpha_0 = \angle P_{1} O P = v (t - t_{\mathrm{prev}}) / R$.

Note that the length of the great circle path from $P_{1}$ to $P$ is $v (t - t_0)$.
Then,
\begin{equation}\label{coordP}
\overrightarrow{OP} \ \ = \ \ \frac{R}{OQ}\overrightarrow{OQ}
\ \ = \ \ \frac{R}{OQ} \left(\overrightarrow{OP_{1}} + \frac{P_{1} Q}{P_{1} P_{2}} \overrightarrow{P_{1} P_{2}}\right).
\end{equation}
By the law of sines in triangle $O P_{1}Q$, using the fact that the angle $\angle O P_{1} Q =(\pi - \alpha) / 2$, it follows that
\begin{equation}
\label{lsine1}
\frac{P_{1} Q}{\sin(\alpha_0)} \ \ = \ \ \frac{OQ}{\sin(\angle O P_{1} Q)}
\ \ = \ \ \frac{OQ}{\sin((\pi - \alpha)/2)}
\ \ = \ \ \frac{OQ}{\cos(\alpha/2)}.
\end{equation}
By the law of sines in triangle $OP_{2}Q$, using the fact that the angle $\angle O P_{2} Q = (\pi - \alpha) / 2$, it follows that
\begin{equation}
\label{lsine2}
\frac{P_{2} Q}{\sin(\alpha - \alpha_0)} \ \ = \ \ \frac{OQ}{\sin(\angle O P_{2} Q)}
\ \ = \ \ \frac{OQ}{\sin((\pi - \alpha)/2)}
\ \ = \ \ \frac{OQ}{\cos(\alpha/2)}.
\end{equation}
Combining \eqref{lsine1}, \eqref{lsine2}, and the relation $P_{1} Q + P_{2} Q = P_{1} P_{2} = 2 R \sin(\alpha/2)$, it follows that
\begin{equation}
\label{firstp1q}
\frac{P_{1} Q}{P_{1} P_{2}} \ \ = \ \ \frac{\sin(\alpha_0)}{\sin(\alpha - \alpha_0) + \sin(\alpha_0)}
\end{equation}
and
\begin{equation}
\label{firstoq}
OQ \ \ = \ \ \frac{P_{1} P_{2} \cos(\alpha/2)}{\sin(\alpha - \alpha_0) + \sin(\alpha_0)}
\ \ = \ \ \frac{2 R \sin(\alpha/2) \cos(\alpha/2)}{\sin(\alpha - \alpha_0) + \sin(\alpha_0)}
\ \ = \ \ \frac{R \sin(\alpha)}{\sin(\alpha - \alpha_0) + \sin(\alpha_0)}.
\end{equation}
Plugging \eqref{firstp1q} and \eqref{firstoq} into \eqref{coordP}, it follows that
\begin{equation}
\label{formP}
\overrightarrow{OP} \ \ = \ \ \frac{1}{\sin(\alpha)} \Big(\sin(\alpha - \alpha_0) \overrightarrow{OP_{1}} + \sin(\alpha_0) \overrightarrow{OP_{2}}\Big).
\end{equation}
Note that, as expected, replacing $\alpha_0$ by 0 (resp. $\alpha$) in \eqref{eqP}, we get $P = P_{1}$ (resp. $P = P_{2}$).

Writing the Cartesian coordinates of $P$ as $P = (x_P,y_P,z_P)$, and substituting $P_{1} = (x_{1},y_{1},z_{1})$ and $P_{2} = (x_{2},y_{2},z_{2})$ into \eqref{formP}, it follows that
\begin{equation}
\label{eqP}
\left( 
\begin{array}{c}
x_P\\
y_P\\
z_P
\end{array}
\right) \ \ = \ \
\frac{1}{\sin(\alpha)} \left(\sin(\alpha - \alpha_0) \left( 
\begin{array}{c}
x_{1}\\
y_{1}\\
z_{1}
\end{array}
\right) + \sin(\alpha_0) \left( 
\begin{array}{c}
x_{2}\\
y_{2}\\
z_{2}
\end{array}
\right)\right),
\end{equation}
and thus it follows that the latitude $\mbox{latitude}(P)$ of $P$ is
\[
\mbox{latitude}(P) \ \ = \ \ \frac{180}{\pi}\arcsin(z_P / R),
\]
and, for $z_P \in (-R,R)$ (if $z_P = R$ or $z_P = -R$ then $P$ is at a pole), the longitude $\mbox{longitude}(P)$ of $P$ is
\[
\mbox{longitude}(P) \ \ = \ \ \frac{180}{\pi} \arccos\left(\frac{x_P}{\sqrt{R^2 - z_P^2}}\right)
\]
for $y_P \geq 0$, and
\[
\mbox{longitude}(P) \ \ = \ \ -\frac{180}{\pi} \arccos\left(\frac{x_P}{\sqrt{R^2 - z_P^2}}\right)
\]
for $y_P < 0$.
The code that implements the computations in this section is available at \url{https://github.com/vguigues/Heuristics_Dynamic_Ambulance_Management}.

\begin{algorithm}
\small
\caption{\textbf{Trajectory computation of ambulance j trips}}
\begin{algorithmic}[1] 
    \State Input: vectors Trips, Types, Times, \(t_{\mathrm{step}}\).
    \State Outputs: discretized\_rides(j), discretized\_ride\_times(j), discretized\_ride\_types(j).
    \State \(t_{\mathrm{next}} \leftarrow \mbox{Times}(j)(2)\)
    \State \(t_{\mathrm{prev}} = \mbox{Times}(j)(1)\)
    \State \(t_{\mathrm{current}} \leftarrow t_{\mathrm{prev}} - (t_{\mathrm{prev}} \mod t_{\mathrm{step}})\)
    \If{\(t_{\mathrm{current}} = t_{\mathrm{prev}}\)}
        \State discretized\_rides(j) \(\leftarrow \) \{Trips(j)(1)\}
        \State discretized\_ride\_types(j) \(\leftarrow \) \{Types(j)(1)\}
        \State discretized\_ride\_times(j) \(\leftarrow \) $\{t_{\mathrm{current}}\}$
    \EndIf
  \State \(i \leftarrow 1\)
  \State $n=|\mbox{Trips(j)}|$
    \While{$i<n$}
        \While{\(t_{\mathrm{current}} + t_{\mathrm{step}} \leq t_{\mathrm{next}}\)}
            \State \(\ell \leftarrow \)  positionBetweenOriginDestination(Trips(j)(i),
            \State $\hspace*{6.7cm}$Trips(j)(i+1),
            \State $\hspace*{6.7cm}$$t_{\mathrm{prev}}, t_{\mathrm{current}} + t_{\mathrm{step}}, v)$             
            \State discretized\_rides(j)\(\leftarrow \) discretized\_rides(j) \(\cup\) \{\(\ell\)\}
            \State discretized\_ride\_times(j) \(\leftarrow \) discretized\_ride\_times(j) \(\leftarrow \)  \(\cup\) $\{t_{\mathrm{current}} + t_{\mathrm{step}}\}$
            \State discretized\_ride\_types(j)\(\leftarrow \) discretized\_ride\_types(j)  \(\cup\) \{Types(j)(i)\}
            \State \(t_{\mathrm{current}} \leftarrow t_{\mathrm{current}} + t_{\mathrm{step}}\)
        \EndWhile
        \If{$(i+1=n)$}\\
           \State $\hspace*{1.2cm}$ $i=i+1$
        \Else{} \\
          \State $\hspace*{1.2cm}$contd=1
         \While{contd}\\
          \State $\hspace*{1.8cm}$i=i+1
              \If{$(i+1>n)$}\\
              \State $\hspace*{2.3cm}$contd=0
              \ElsIf{($t_{\mathrm{current}} + t_{\mathrm{step}} \leq \mbox{Times}(j)(i+1)$)}\\
              \State $\hspace*{2.3cm}$contd=0
              \EndIf  
         \EndWhile
        \If{($i<n$)}
        \State \(t_{prev} \leftarrow \) Times(j)(i)
        \State \(t_{next} \leftarrow \) Times(j)(i+1)
       \EndIf    
    \EndIf
    \EndWhile
\end{algorithmic}
\label{algo_discretization}
\end{algorithm}

\subsection{Visualization of performance metrics}
\label{sec:vis}

The output of the simulation contains performance metric data.
The two main types of performance metrics are response times and penalized responses as specified by the penalization function {\tt{penalization}} given in \eqref{penbtct}.
The link {\em{Distribution of Response Times}} available clicking the link 
{\em{SAMU-Response Times}} 
allows us to analyze the response times to pick up patients
for the allocation policies described in
Section \ref{sec:ambmodel}.
Three visualization tools are provided for the distribution
of response times for emergency calls: Cumulative Distribution Function (CDF), tables providing some statistical
indicators, and histograms of the distributions.
For all these visualizations, three filters need to be chosen: the day(s)
of the week, a selector indicating if the output is a real or penalized
response time (for penalized waiting time the penalization function {\tt{penalization}} given in \eqref{penbtct} is used with
$\theta_c=1$ for low priority calls, $\theta_c=2$ for intermediate priority calls,  and
$\theta_c=4$ for high-priority calls), and time windows of 30 mins.
We then have three selectors
providing these
performance metrics
can be visualized in three ways: 
(1)~cumulative distribution function of the performance metric, (2)~a table that provides summary statistics of the performance metric, and (3)~a histogram of the performance metric distribution.
For each of these visualizations, three filters are specified: (a)~the day(s) of the week, (b)~the time range within each day, and (c)~a selector of the performance metric (response times or penalized responses) for which to display results.

Below we show examples of the visualization of the penalized response performance metric, with the parameter $\theta_{c}$ in~\eqref{penbtct} chosen as follows: $\theta_{c} = 1$ for low priority emergencies, $\theta_{c} = 2$ for intermediate priority emergencies, and $\theta_{c} = 4$ for high-priority emergencies.

\subsubsection{Cumulative distribution function}

\begin{figure}
\centering
\begin{tabular}{c}
\includegraphics[width=\textwidth]{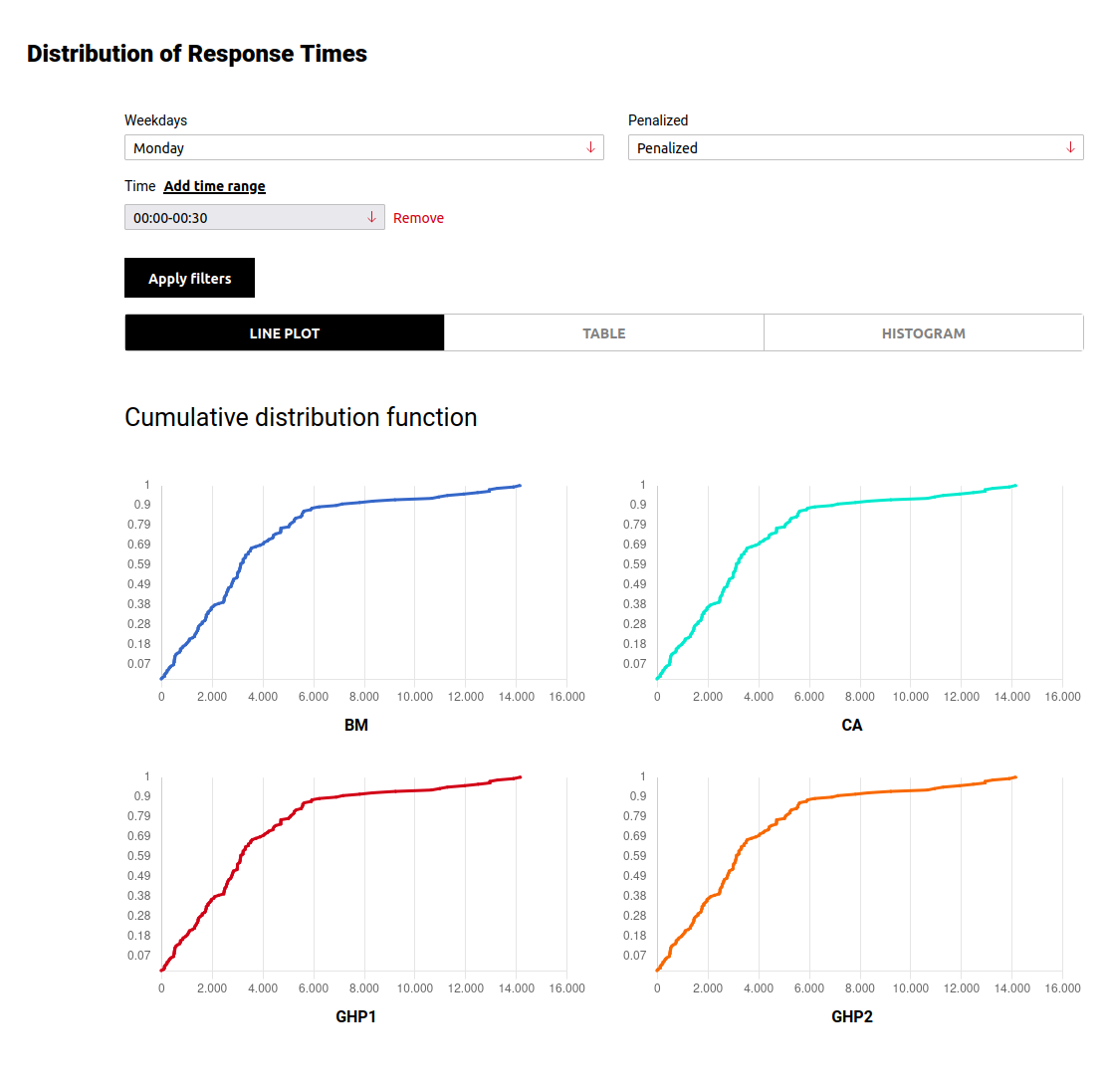}
\end{tabular}
\caption{Cumulative distribution functions of the penalized response performance metric under different ambulance dispatch policies.}\label{fig10}
\end{figure}

The {\em{Line plot}} selector  
available from
the link {\em{Distribution of Response Times}} provides
the cumulative distribution functions of response times. 
An example of the cumulative distribution functions of the penalized response performance metric under five ambulance dispatch policies are shown in Figure~\ref{fig10}.

\subsubsection{Tables}

The \emph{Table} selector 
available from 
the link {\em{Distribution of Response Times}} 
displays, for each ambulance dispatch policy, the maximum, minimum, mean, and 0.9 quantile of the distribution of the chosen performance metric.

\if{
\begin{figure}
\centering
\begin{tabular}{c}
\includegraphics[scale=0.15]{tableresponsetime.png}
\end{tabular}
\caption{Mean, min, max, and 0.9-quantile of the distribution of response time with five allocation policies.}\label{fig11}
\end{figure}
}\fi

\subsubsection{Histograms}

\begin{figure}
\centering
\begin{tabular}{c}
\includegraphics[width=\textwidth]{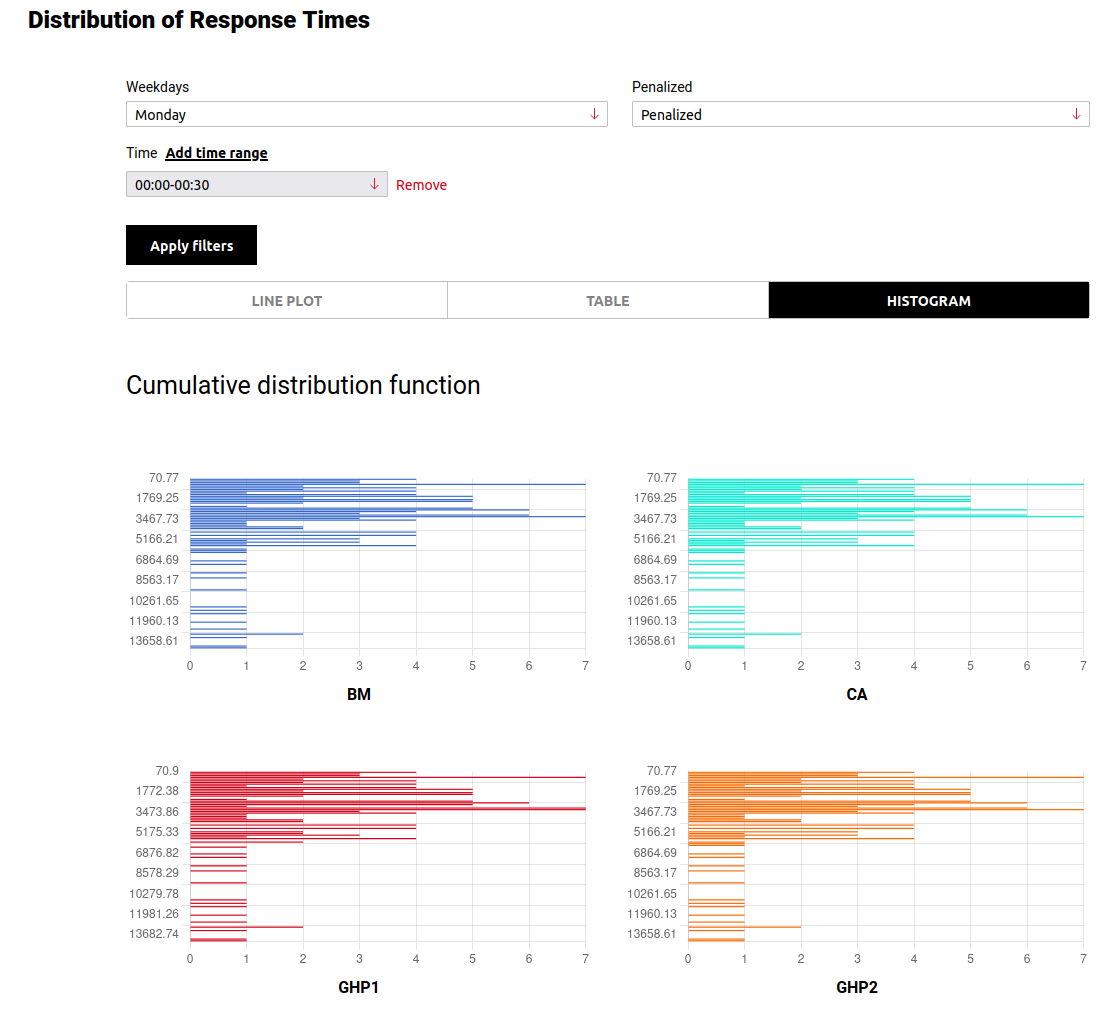}
\end{tabular}
\caption{Histograms of the penalized response performance metric under different ambulance dispatch policies.}\label{fig12}
\end{figure}

The \emph{Histogram} selector 
available from
the link {\em{Distribution of Response Times}} 
provides histograms
of the distrbution of
penalized response time.
An example of histograms of the penalized response performance metric under five ambulance dispatch policies are shown in Figure~\ref{fig12}.

\subsection{Running the ambulance dispatch heuristics with code available on GitHub}
\label{sec:codeh}

The subdirectory SAMU3 of the GitHub repository contains the implementations of the heuristics described in section~\ref{sec:heur}, considering the graph of the streets of Rio de Janeiro. The implementations can be found in the source file src/solver.cpp. A class was built for every heuristic. These classes are QueueSolver for CA heuristics, 
ForwardSolver for BM heuristic, PrioritySolver for GHP1 heuristic, and MinMaxSolver for GHP2 heuristic.
The code is written in C++ and has the following dependencies:

\begin{itemize}
    \item Boost: available at \url{https://www.boost.org/} or via Linux APT package manager, used for file system functions and configuration files.
    \item fmtlib: available at \url{https://github.com/fmtlib/fmt}, used for better and faster printing.
    \item xtl: available at \url{https://github.com/xtensor-stack/xtl}, used for multidimensional arrays.
    \item xtensor: available at \url{https://github.com/xtensor-stack/xtensor}, also used for multidimensional arrays.
\end{itemize}

The fmtlib, xtl and xtensor libraries  are installed via CMake. The general procedure involves downloading each library
and creating a build directory
in the library root directory:
\begin{verbatim}
    unzip xtensor-master.zip
    cd xtensor-master
    mkdir build
\end{verbatim}

Once the build directory is created, we run the following commands:

\begin{verbatim}
    cd build
    cmake ..
    make
    sudo make install
\end{verbatim}

Once all dependencies are installed, the Makefile can be used to compile the code. From the source code repository, run:

\begin{verbatim}
    cd SAMU3
    make
\end{verbatim}

The make command will generate an executable called esma. To run the heuristics, a configuration file
must be provided with a -f option:

\begin{verbatim}
    ./esma -f test.cfg
\end{verbatim}

The file test.cfg in SAMU3 directory gives examples of some parameters that can be modified, among them:

\begin{itemize}
    \item h\_use\_fixed\_bases: flag that indicates whether ambulances should return to a fixed base or to the closest base after finishing  service.
    \item n\_scenarios: number of scenarios (weeks) in the simulation of the heuristics.
    \item n\_hospitals: number of hospitals (maximum of nine)
    \item n\_bases: number of bases (maximum of 29)
    \item n\_ambulances: number of ambulances in the simulation.
    \item output\_folder: directory where the trajectories and scenarios will be stored.
\end{itemize}

The parameters of the configuration file may also be passed via the command line. Note that if the user passes a parameter in the command line, the value in the configuration file is ignored and the command line argument is used instead.

By running the above command, the simulation will be performed for each of the five heuristics described in section~\ref{sec:heur} for the number of weeks given in n\_scenarios. The results are saved in the path given by output\_folder. 

The calls generated are saved in calls.txt file. 

A subdirectory that saves trajectories and response times is created for each heuristic. 

{\textbf{Trajectories files.}} The trajectory files are named "output\_scenarios\_id\_h", where id is the scenario index and h is the name of the heuristic. Each line contains the list of active calls, followed by a description of the active calls (call id, call location, call priority) followed by the number of ambulances and a description of the current state of each ambulance. The description of each ambulance is given by its id, its ride type (see Section \ref{sec:discamb}), its current location and the index of its current destination: either a call id, a hospital id, or a base id, depending on the ambulance ride type. 

{\textbf{Response times files.}} The response times files are named after each heuristic and contain for each call: the call id, the instant of the call, the response time, the allocation cost, and the index of the ambulance that served the call.

\section{Implementation Details}
\label{sec:tech_stack}

The website was developed using Python, Django, PostgreSQL as database, Redis 6.2 as cache database, and Celery message broker as backend technologies for their overall capacity to consume data, manipulate it, and serve it on a low-latency online server capable of receiving requests from the data visualization application. 
Our project interacts with the user through filters with which the user can select various alternatives.
For our project, we used a technology with a powerful initial renderer and high re-rendering capacity from interactions with the user.
More precisely, React was chosen as the base framework of the project, react-chartjs-2 3.3 library was used for charts, and React Leaflet for the visualizations on maps.

\if{
\subsection{App report section}
The app contains two models: Ambulance call and Upload. The ambulance call model contains latitude and longitude, accident time, hospital time, ambulance calls type and priority data. The Upload model represents a dataset uploaded by a regular user or an admin at the admin panel. The file sent to our project needs to contain a list of ambulance calls and, in each of them, the location information of the emergency call and the information of the patient who needs medical attention, such as name, email, and organization name.

}\fi

\if{

\subsection {Visualization}
The page header holds options for selecting a database used in the visualization and an option for downloading that data. On the right side of the header the website navigation.
There is an informative banner at the centre of the page explaining how to use the tool and an “upload data” action. Below is a tab-based carousel of charts with data from Rio de Janeiro’s 2021 ambulance call history.

Four charts are available: line plots, heatmaps, bar charts and pie charts. Each one holds standard filters, such as the date and time ranges, and specific ones, such as age, gender, the call type and priority. Below each chart, there is an option to download the filtered data.

Returning to the top banner, more specifically to the “upload data” call-to-action, a pop-up form allows any user to create a new Upload instance. This database will be available for 24h via a unique and shareable link.
Each new Upload instance requires a text file which, when processed, is used to create Call instances linked to it. This process is done asynchronously in a worker parallel to the backend server.
When added via the admin panel, the Upload will be persistent and available for any user as a “database” option on the data-visualization webpage as it occurs with the Rio de Janeiro database.
}\fi

\section{Conclusion}
\label{sec:conclusion}

This paper describes a website that can be used to simulate and visualize ambulance operations in a chosen region.
Such a tool should be useful to emergency medical services and researchers working on related problems.
Specifically, data (real or synthetic) of an EMS system as well as emergencies associated with a specified time period can be uploaded, different views of the data can be displayed, forecasts of emergency arrival rates can be estimated using the data, different views of the forecasts can be displayed, the forecasts can be used to generate random emergencies in a simulation, ambulance operations can be simulated under specified ambulance dispatch policies, the ambulance movements can be viewed on a map, and performance metrics can be computed and displayed.
The ambulance dispatch policies provided on the website are described in \cite{ourheuristics23a}.
Future enhancements include the ability of users to plug in code for alternative ambulance dispatch policies, to generate and visualize results for these policies.

\addcontentsline{toc}{section}{References}
\bibliographystyle{plain}
\bibliography{Biblio}

\end{document}